\documentclass[%
 reprint,
 amsmath,amssymb,
 aps,
 pra,
]{revtex4-2}
\usepackage[ruled,linesnumbered]{algorithm2e}
\let\oldnl\nl
\newcommand{\nonl}{\renewcommand{\nl}{\let\nl\oldnl}}
\usepackage{ragged2e}
\usepackage{graphicx}
\usepackage{dcolumn}
\usepackage{bm}
\usepackage{amsthm}

\usepackage{booktabs}
\usepackage{multirow}

\begin{document}

\preprint{APS/123-QED}

\title{Quantum Multi-view Kernel Learning with Local Information}
\author{Jing Li$^{1,2,3}$}{}
\author{Yanqi Song$^{4}$}{}
\author{Sujuan Qin$^{1}$}{}
\author{Fei Gao$^{1}$}\email{gaof@bupt.edu.cn}

\affiliation{$^{1}$State Key Laboratory of Networking and Switching Technology, Beijing University of Posts and Telecommunications, Beijing,100876, China}

\affiliation{$^{2}$National Engineering Research Center of Disaster Backup and Recovery,
Beijing University of Posts and Telecommunications, Beijing, 100876, China}
\affiliation{$^{3}$School of Cyberspace Security, Beijing University of Posts and Telecommunications, Beijing, 100876, China}
\affiliation{$^{4}$China Academy of Information and Communications Technology, Beijing,100191, China}

\date{\today}

\begin{abstract}
Kernel methods serve as powerful tools to capture nonlinear patterns behind data in machine learning. The quantum kernel, integrating kernel theory with quantum computing, has attracted widespread attention. However, existing studies encounter performance bottlenecks when processing complex data with localized structural patterns, stemming from the limitation in single-view feature representation and the exclusive reliance on global data structure. In this paper, we propose quantum multi-view kernel learning with local information, called L-QMVKL. Specifically, based on the multi-kernel learning, a representative method for multi-view data processing, we construct the quantum multi-kernel that combines view-specific quantum kernels to effectively fuse cross-view information. Further leveraging local information to capture intrinsic structural information, we design a sequential training strategy for the quantum circuit parameters and weight coefficients with the use of the hybrid global-local kernel alignment.
We evaluate the effectiveness of L-QMVKL through comprehensive numerical simulations on the Mfeat dataset, demonstrating significant accuracy improvements achieved through leveraging multi-view methodology and local information. Meanwhile, the results show that L-QMVKL exhibits a higher accuracy than its classical counterpart. Our work holds promise for advancing the theoretical and practical understanding of quantum kernel methods.

\end{abstract}

\maketitle


\section{Introduction}\label{section1}
\par Machine learning has achieved remarkable progress in recent decades, giving rise to a variety of multidisciplinary implementations widely used in diverse industries~\cite{lecun2015deep,jordan2015machine}. However, effectively handling complex non-linear data remains a persistent challenge. In response, the kernel theory~\cite{hofmann2008kernel,gonen2011multiple} has emerged as an elegant solution, employing kernel functions to implicitly compute inner products in high-dimensional feature spaces without requiring explicit feature mapping. Consequently, kernel methods have garnered significant attention and become indispensable tools in the field of machine learning.

\par Concurrently, escalating data complexity and computational demands have motivated the exploration of quantum computing, a paradigm offering provable speedups for classically challenging tasks~\cite{shor1994algorithms,grover1997quantum, harrow2009quantum}. The convergence between quantum computing and machine learning has spawned Quantum Machine Learning (QML)~\cite{biamonte2017quantum,dunjko2018machine,houssein2022machine,huang2023near}, which has rapidly gained prominence and demonstrated notable advancements in both supervised~\cite{rebentrost2014quantum,schuld2020circuit,abbas2021power,beer2020training,song2024quantum,wu2025fuzzy} and unsupervised~\cite{bondarenko2020quantum,kerenidis2021quantum,li2025efficient,wu2025resource} learning.

\begin{figure*}
 \centering
 \includegraphics[width = 14.25 cm]{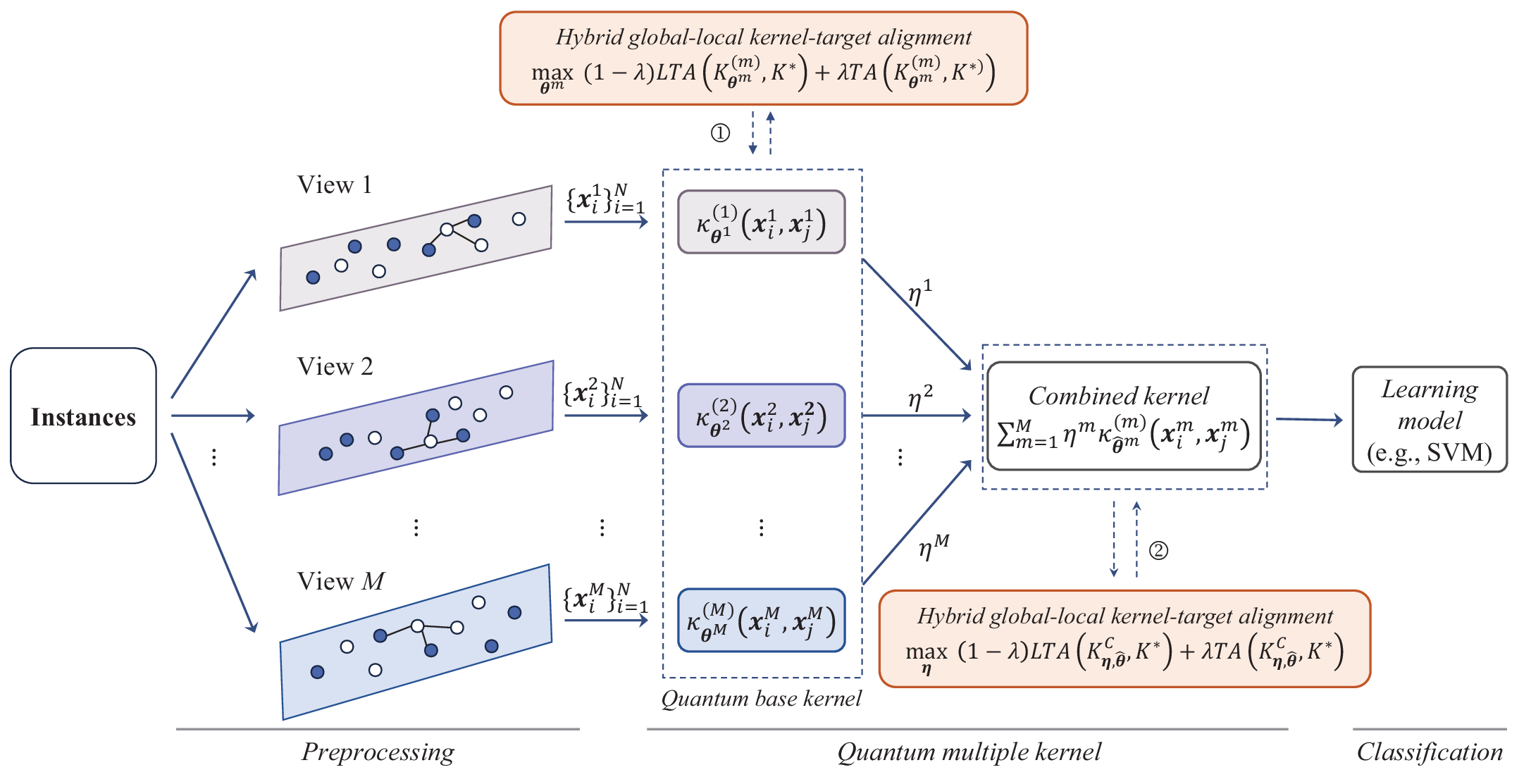}
 \caption{Work flow of L-QMVKL.}
 \label{fig:flow}
\end{figure*} 

\par As one of the representative methods of QML, quantum kernel~\cite{havlivcek2019supervised,schuld2019quantum,schuld2021supervised} is a quantum extension of the classical kernel theory. Relevant algorithms utilize quantum computing devices to evaluate kernel functions that correspond to data-encoding feature maps defined as transformations embedding classical data into quantum Hilbert spaces. 
Current research has yielded diverse quantum kernel architectures, such as the quantum evolution kernel~\cite{henry2021quantum} and the quantum neural tangent kernel~\cite{liu2022representation,incudini2023quantum, shirai2024quantum}, with their trainability and related characteristics being intensively investigated~\cite{liu2021rigorous,wang2021towards,hubregtsen2022training,thanasilp2024exponential}.
Quantum multi-kernel stands out as a special form. By combining a set of quantum kernels (called quantum base kernels), quantum multi-kernel is able to effectively exploit the complementary information across diverse kernels. Specifically, Miyabe et al. ~\cite{miyabe2023quantum} and Fu et al.~\cite{fu2024exploiting} leveraged fixed-parameter quantum kernels to construct quantum multi-kernels, with their approaches demonstrating promising performance on the German Credit dataset and Google Commands dataset, respectively. Further enhancing the flexibility, Vedaie et al. ~\cite{vedaie2020quantum} and Ghukasyan et al.\cite{ghukasyan2023quantum} employed trainable quantum kernels based on Parameterized Quantum Circuits (PQCs).
Nevertheless, these approaches remain limited by single-view and homogeneous data representations derived from a single source or modality. This limitation stands in contrast to real-world scenarios where data inherently consists of heterogeneous feature representations spanning multiple views and modalities. Moreover, the parameter training process in conventional quantum kernel frameworks predominantly depends on global information, intensifying the difficulties in capturing intrinsic local structure. These dual constraints adversely affect the discriminative power of current quantum kernel models in handling complex data in the real world. 

In classical machine learning, multi-view learning methodology and local information-based training strategy provide effective solutions to enhance the model's ability to process complex data. Specifically, multi-view learning~\cite{zhao2017multi,yu2025review} exploits complementary information across diverse data representations from various sources or modalities, enabling the enhancement of the expressive power of models. A natural approach involves constructing the multi-kernel through the combination of view-specific kernels, which has demonstrated proven effectiveness in multi-view feature representation and fusion~\cite{yeh2012novel,chen2014emotion,yan2023towards}. The local information-based training strategy has been developed to improve the model's performance by leveraging the local structure of data. One representative in kernel learning is local kernel alignment maximization~\cite{wang2017multiple}, which has been applied in multi-kernel clustering~\cite{wang2018local,zhang2021late}.
A question arises naturally: Whether the two approaches are effective in overcoming the aforementioned limitations of quantum kernels? In response to this question, we investigate the synergistic integration between quantum kernels and multi-view learning with the local information-based training strategy.

\par In this paper, we propose a novel quantum multi-view kernel learning with local information, called L-QMVKL. In L-QMVKL, the quantum multiple-kernel is constructed by combining view-specific quantum base kernels that are optimized quantum kernels for different views. Particularly, for parameter training, we present a sequential training strategy leveraging local information. With the objective of hybrid global-local kernel-target alignment maximization, the circuit parameters of quantum base kernels are learned first to capture the intrinsic features in different views of the dataset. Subsequently, the combination weight coefficients of the quantum multi-kernel are trained to fuse the cross-view features. By exploiting complementary information across diverse views and local structures of data, the proposed L-QMVKL exhibits effective characterization and fusion of multi-view data features. 

\par To validate the effectiveness of L-QMVKL, we conduct extensive numerical simulations on the Mfeat dataset comprising 6 views. Results show that L-QMVKL achieves significant accuracy improvements over both the single-view models and global information-based methods. Besides, we compare our algorithm to its classical counterpart, demonstrating the superior accuracy as well. Specifically, our experimental analysis reveals four key observations:
\begin{itemize}
\item  When the multi-view learning methodology is independently implemented, the model demonstrates superior performance over single-view quantum models, achieving a minimum accuracy improvement of 9.5\% at $P = 6$.

\item When the local information-based training strategy is independently implemented, the model outperforms global information-based approaches on three key features (FOU, ZER, and MOR), achieving peak accuracy gains of 3.81\% at $k=8$ and $P = 6$.

\item The proposed L-QMVKL, which integrates both approaches, exhibits at least 10.06\% enhanced performance compared to single-view models based on global information at $\lambda = 0.25$, $k=8$, and $P = 6$. 

\item The proposed L-QMVKL surpasses classical counterparts by 1.44\% in the average accuracy at $\lambda = 0.125$ and $k=8$.
\end{itemize}
Here, $\lambda$ and $k$ denote the hybrid parameter and the number of neighbors, respectively, in the hybrid global-local kernel alignment, while $P$ represents the depth of the quantum circuit. 

\par The rest of this paper is organized as follows. We propose the structure and the training of L-QMVKL in Section~\ref{sec:2} and Section~\ref{sec:3}, respectively.
The results of numerical simulations are presented in Section~\ref{sec:4}. Finally, we summarize the paper and present future works in Section~\ref{sec:6}.

\section{Structure of L-QMVKL}
\label{sec:2}
\par As shown in~FIG.~\ref{fig:flow}, L-QMVKL consists of 3 components: preprocessing, quantum multi-kernel, and classification. Among them, the quantum multi-kernel is the core of L-QMVKL, which combines view-specific quantum kernels to harness the comprehensive information from diverse views. 

\par To construct the view-specific quantum kernel that can be suitable for each view, we start by selecting base kernels from a set of trainable quantum kernels, enabling them to be adaptable for heterogeneous data representations. Then, each quantum base kernel is optimized to effectively capture the nonlinear structure of each view. Here, trainable quantum kernels can be implemented by a parameterized operation $W(x,\theta)$ composed of PQC and the data-encoding block. Given an initial quantum state $|\psi_0\rangle$, the corresponding data-encoding feature map $\phi:\mathcal{X}\to \mathcal{H}$ and quantum kernel $\kappa: \mathcal{X} \times \mathcal{X} \rightarrow \mathbb{R}$ are defined as
\begin{align}
\phi(\bm{x}_i) &= W(\bm{x}_i,\theta)|\psi_0\rangle \langle \psi_0| W^{\dagger}(\bm{x}_i,\theta),\\
 \kappa(\bm{x}_i,\bm{x}_j)& = \text{Tr}[\phi(\bm{x}_j)\phi(\bm{x}_i)] \\
 &= {| \langle \psi_0 | W^{\dagger} (\bm{x}_j,\bm{\theta})W (\bm{x}_i,\bm{\theta}) |\psi_0\rangle|} ^2, 
\end{align}
where $\mathcal{X}$ and $\mathcal{H}$ represent the original input space and the quantum Hilbert space, respectively. Thus, by predefining a set of parameterized operations $\mathcal{W}=\{W_1, W_2,\cdots\}$, diverse quantum kernels can be formed. In practice, $\mathcal{W}$ can be configured according to quantum devices (e.g., Hardware Efficient Ansatz (HEA)~\cite{kandala2017hardware}) or specific problems (e.g., Quantum Alternating Operator Ansatz (QAOA)~\cite{farhi2014quantum,hadfield2019quantum}).

\par Specifically, given a multi-view dataset comprising $M$ views, through preprocessing, the extracted feature vectors are $\{ \bm{x}_i^{1}\}_{i=1}^N, \cdots, \{ \bm{x}_i^{M}\}_{i=1}^N$ where $\bm{x}_i^{m} \in \mathbb{R}^{d_m}$. For the $m$-th view, we select $W^m \in \mathcal{W}$ and let the initial state be $|0\rangle$, then the corresponding quantum base kernel is expressed as
\begin{equation}
\kappa_{\bm{\theta}^m}^m (\bm{x}_i^m,\bm{x}_j^m) = {| \langle 0 | (W^m)^{\dagger} (\bm{x}_j^m,\bm{\theta}^m)W^m (\bm{x}_i^m,\bm{\theta}^m) |0\rangle|} ^2.
\label{eq:qk}
\end{equation}

\par After constructing the set of quantum base kernels $\{ \kappa_{\bm{\theta}^m}^m \}_{m=1}^M$, given the combination function $f_{\bm{\eta}}$ with weight coefficients $\bm{\eta}$, the quantum multi-kernel is defined by
\begin{equation}
 \kappa^c_{\bm{\theta},\bm{\eta}} (\bm{x}_i,\bm{x}_j) = f_{\bm{\eta}} (\{ \kappa_{\bm{\theta}^m}^m (\bm{x}_i^m,\bm{x}_j^m) \}_{m=1}^M),
\end{equation}
where $\bm{\theta}=\{ \bm{\theta}^1,\cdots, \bm{\theta}^M \}$ denotes the parameters of base kernels. Considering a linear combination function $f_{\bm{\eta}} = \langle \bm{\eta}, \cdot \rangle$, the linear combined kernel is
\begin{equation}
 \kappa^c_{\bm{\theta},\bm{\eta}} (\bm{x}_i,\bm{x}_j) = \sum_{m=1}^M \eta^m \kappa_{\bm{\theta}^m}^m (\bm{x}_i^m,\bm{x}_j^m),
\end{equation}
where $\bm{\eta}\in \mathbb{R}_{+}^{M}$ and $\sum_{m=1}^M \eta^m = 1$. The optimized quantum multi-kernel serves as input to a kernel-based classifier, such as the Support Vector Machine (SVM)~\cite{suykens1999least}.

\section{Training of L-QMVKL}

\label{sec:3}

\subsection{Cost function}
\par In kernel learning, kernel alignment~\cite{wang2015overview} is an important tool for evaluating the similarity between two kernels for a given dataset. Given two kernels $\kappa_1$ and $\kappa_2$ with corresponding kernel matrices $K_1$ and $K_2$, kernel alignment is defined as
\begin{equation}
 A (K_1,K_2)=\frac{{\langle K_1, K_2\rangle}_F}{{\langle K_1, K_1\rangle}_F{\langle K_2, K_2\rangle}_F},
\end{equation}
where, ${\langle K_1, K_2\rangle}_F = \sum_{i=1}^{N} \sum_{j=1}^N \kappa_1 (\bm{x}_i,\bm{x}_j) \kappa_2 (\bm{x}_i,\bm{x}_j)$. For supervised learning, each point $\bm{x}_i$ is associated with a label $y_i$. An ideal target kernel can be defined as 
\begin{equation}
 \kappa^{*} =\left\{\begin{matrix}
 1& y_i = y_j\\
 -1 & y_i \neq y_j
\end{matrix}\right. .
\end{equation}
In the case that $y_i \in \{-1,1\}$, i.e. binary classification, the target kernel matrix is equal to $K^{*}=\bm{y} \bm{y}^{T}$, where $\bm{y} = (y_1,\cdots,y_N)^{T}$. To assess the similarity between a kernel $\kappa$ and the ideal target kernel $\kappa^{*}$, kernel-target alignment is
\begin{equation}
 TA (K, K^{*}) = \frac{{\langle K, K^{*}\rangle}_F}{{\langle K, K\rangle}_F{\langle K^{*}, K^{*} \rangle}_F} =\frac{\sum_{ij} y_iy_j\kappa (\bm{x}_i,\bm{x}_j)}{N\sqrt{\sum_{ij}{\kappa (\bm{x}_i,\bm{x}_j)}^2}} . 
\end{equation}

\par Further leveraging the local information for the given dataset, the local kernel-target alignment~\cite{wang2017multiple} is formed as
\begin{equation}
LTA (K, K^{*}) = \frac{1}{N} \sum_{i}^{N} TA (K_i, K_i^{*}),
\end{equation}
where $K_i$ represents the $i$-th local kernel matrix defined by

\begin{equation}
[K_i]_{j,h} = \kappa (\bm{x}_j, \bm{x}_h), \quad \bm{x}_j, \bm{x}_h \in \mathcal{N} (\bm{x}_i),
\end{equation}
and $\mathcal{N} (\bm{x}_i)$ denotes the set of $k$ nearest neighbors of $\bm{x}_i$. The local target matrix $K_i^{*}$ is given by $\bm{y}_i \bm{y}_i^{T}$, where the elements of $\bm{y}_i$ correspond to the labels of the $k$ nearest neighbors of $\bm{x}_i$.

\par Combining global and local kernel-target alignment, the hybrid global-local kernel-target alignment is
\begin{equation}
 HTA (K, K^{*}) = (1-\lambda) LTA (K, K^{*}) + \lambda TA (K, K^{*}),
\end{equation}
where the hybrid parameter $\lambda \in [0,1]$.

\subsection{Optimization} 
\par The model optimization consists of two sequential stages. Firstly, circuit parameters $\bm{\theta}^m$ of the quantum base kernel are trained to capture the intrinsic structure of data within each view. Subsequently, the weight coefficients $\bm{\eta}$ of the quantum multi-kernel are learned to effectively achieve cross-view feature fusion. The detailed process is illustrated below.

\textbf{Stage 1}: Training $\bm{\theta}^m$ of quantum base kernel for each view.
\par The quantum base kernel is trained
with the objective of maximizing the hybrid global-local kernel-target alignment. For the $m$-th view, the cost function is
\begin{equation}
 \arg \max_{\bm{\theta}^m} HTA (K^{m}_{\bm{\theta}^m}, K^{*}).
\end{equation}
Here, since the $k_1$ nearest neighbors for each data point are required in the computation of $LTA$, $M$ sets of nearest neighbors $\mathcal{N}^m_i$ are constructed initially for the various features. Then, we utilize the gradient descent method~\cite{ruder2016overview} as the classical optimizer. Concretely, for the global kernel-target alignment, its analytic gradient is
\begin{widetext}
\begin{equation}
 \bm{g}_{TA} = \frac{\sum_{ij}y_iy_j \nabla \kappa_{
 \bm{\theta}^m} (\bm{x}_i^m,\bm{x}_j^m) \sum_{ij} \kappa_{
 \bm{\theta}^m} (\bm{x}_i^m,\bm{x}_j^m)^2 }{ (\sum_{ij} \kappa_{
 \bm{\theta}^m} (\bm{x}_i^m,\bm{x}_j^m)^2)^{\frac{3}{2}}} -\frac{\sum_{ij} y_i y_j\kappa_{
 \bm{\theta}^m} (\bm{x}_i^m,\bm{x}_j^m) \sum_{ij}\kappa_{
 \bm{\theta}^m} (\bm{x}_i^m,\bm{x}_j^m) \nabla \kappa_{
 \bm{\theta}^m} (\bm{x}_i^m,\bm{x}_j^m)}{ (\sum_{ij} \kappa_{
 \bm{\theta}^m} (\bm{x}_i^m,\bm{x}_j^m)^2)^{\frac{3}{2}}},
 \label{eq:gta}
\end{equation} 
\end{widetext}
where $\nabla \kappa_{\bm{\theta}^m} (\bm{x}_i^m,\bm{x}_j^m)$ can be compute by the parameter shift rule~\cite{romero18strategies, schuld2019evaluating}. Then, the gradients of local kernel-target alignments for each point $\bm{g}_{TA}^{i}$ can be computed in the same way. The total gradient is 
\begin{equation}
 \bm{g}_{HTA} = \frac{1-\lambda}{N} \sum_{i=1}^{N} \bm{g}_{TA}^{i} +\lambda \bm{g}_{TA}. 
 \label{eq:ghta}
\end{equation}
The parameters are updated by 
\begin{equation}
 \bm{\theta}^m_{t+1} \leftarrow \bm{\theta}^m_{t} + \xi \bm{g}_{HTA}(\bm{\theta}^m_{t}),
\end{equation}
where $\xi$ represents the learning rate.
\par For each view, update the quantum base kernel parameters until a predefined number of iterations is reached. We denote the optimized parameters as $\hat{\bm{\theta}}=\{ \hat{\bm{\theta}}^1,\cdots,\hat{\bm{\theta}}^M \}$.

\textbf{Stage 2}: Training $\bm{\eta}$ of the quantum multi-kernel.
\par The combined weights $\bm{\eta}$ are learned based on the hybrid global-local kernel-target alignment as well. The optimization object is
\begin{align}
\arg \max_{\bm{\eta}}\textit{ } & HTA (K^c_{\bm{\eta},\hat{\bm{\theta}}},K^{*}), \\
 s.t. \textit{ }& \eta^m>0, m = 1, \cdots, M; \\
& \sum_{m=1}^M \eta^m = 1.
\end{align}
Here,
\begin{widetext}
\begin{equation}
HTA (K^c_{\bm{\eta},\hat{\bm{\theta}}},K^{*}) = \frac{1-\lambda}{Nk} \sum_{i=1}^N \frac{\langle \sum_{m=1}^M \eta^m K_i^m,K_i^{*} \rangle_F}{\sqrt{\langle\sum_{m=1}^M \eta^m K_i^m,\sum_{m=1}^M \eta^m K_i^m \rangle_F}} + \frac{\lambda}{N} \frac{\langle \sum_{m=1}^M \eta^m K^m,K^{*} \rangle_F}{\sqrt{\langle\sum_{m=1}^M \eta^m K^m,\sum_{m=1}^M \eta^m K^m \rangle_F}}.
\end{equation}
\end{widetext}
Notably, the construction of the local combined kernel matrix $\sum_{m=1}^M \bm{\eta}^mK_i^m$ and local target matrix $K_i^{*}$  requires determining the set of nearest neighbors $\mathcal{N}^C_i$ for the $i$-th instance. However, direct construction of $\mathcal{N}^C_i$ presents challenges due to view-specific variations in nearest neighbor sets $\mathcal{N}^m_i$ across different input spaces. To resolve this inherent inconsistency, we select the $k_2$ instances exhibiting the largest values of the quantum multi-kernel similarity measure $\kappa^c_{\bm{\eta},\hat{\bm{\theta}}} (\bm{x}_i,\cdot)$ between the $i$-th instance and other data points. This selection criterion ensures that 
$\mathcal{N}^C_i$ effectively preserves the intrinsic feature space structure associated with the optimized quantum base kernels. Notably, the neighbor set $\mathcal{N}^C_i$ is dynamically updated during iterative optimization of $\bm{\eta}$, thereby maintaining local alignment between the quantum multi-kernel and the target kernel throughout the learning process.

Subsequently, define additional variables $\tau_i = \frac{\bm{\eta}^T \mathcal{M}^i\bm{\eta}}{\bm{\eta}^T \mathcal{M}\bm{\eta}}$ where $\mathcal{M}_{qr}^i= \langle K_i^{q},K_i^{r} \rangle_F$ and $\mathcal{M}_{qr}= \langle K^{q},K^{r}\rangle_F$, and let $a_q = \frac{1}{Nk}\sum_{i=1}^N \frac{1}{\sqrt{\tau_i}} \langle K^{q}_i,K^{*}_i\rangle_F$, $b_q = \frac{1}{N} \langle K^{q},K^{*}\rangle_F$, then the objective function can be rewritten as
\begin{equation}
 \arg \max_{\bm{\eta}} { \frac{ (1-\lambda)\bm{\eta}^T\bm{a}+\lambda \bm{\eta}^T\bm{b}}{\sqrt{\bm{\eta}^T \mathcal{M}\bm{\eta}}} }.
\end{equation}
Fixing $\bm{\tau}$, the weight coefficients $\bm{\eta}$ can be obtained by solving the quadratic programming as follows,
\begin{equation}
 \arg \min_{\bm{\mu}>\bm{0}} {\bm{\mu}^T \mathcal{M} \bm{\mu}} - (1-\lambda)\bm{\mu}^T\bm{a}-\lambda\bm{\mu}^T\bm{b}.
 \label{eq:2ob}
\end{equation}
Then, $\eta^m = \mu_m/\sum_m \mu_m$. By solving $\bm{\eta}$ and $\bm{\tau}$ alternately with the update of $\mathcal{N}^C_i$, the optimized $\bm{\eta}$ is obtained.
\par Overall, the entire training process of L-QMVKL is outlined in Algorithm~\ref{al:1}.

\begin{algorithm}[htbp]
\label{al:1}
\RaggedRight
\caption{Training process of L-QMVKL}
 \nonl \textbf{Stage 1:} Training circuit parameters of quantum base kernels \\
 \KwIn {$\{\bm{X}^m\}_{m=1}^{M}$, $\bm{y}$, $\mathcal{W}$, $M$, $\lambda_0$, $k_1$, $\epsilon_1$, and $T_1$. }
 \KwOut{ ${{\{{{K}}}^m}\}_{m=1}^M$, and $\hat{\bm{\theta}}$.} 
 
 According to $\bm{y}$, calculate the target kernel matrix $K^{*}$; \\
 
\For{$m =1,2,\cdots,M$}{
 Select $W^m \in \mathcal{W}$, and randomly generate initial parameters $\bm{\theta}_{\text{random}}^m$;\\
 Initialize $\bm{\theta}_1^m=\bm{\theta}_{\text{random}}$, ${HTA}_0^m=0$, and $t=1$;\\
 Compute pairwise distances and obtain the $k_1$ nearest neighbors $\mathcal{N}^m_i$ for each instance;\\
 According to $\{\mathcal{N}^m_i\}_{i=1}^N$ and corresponding class labels, construct $K_i^{*}$;\\
 \Repeat {$\left\|{HTA}^m_t-{HTA}^m_{t-1} \right\|\le\epsilon_1$ or $t=T_1$} {
 Compute $\{\kappa_{
 \bm{\theta}^m} (\bm{x}_i^m,\bm{x}_j^m)\}_{i,j=1}^{N}$ according to Eq.~(\ref{eq:qk}) on the quantum computing device, and construct ${{{K}}}^m(\bm{\theta}_t^m)$;\\
 According to $\mathcal{N}^m_i$, construct ${{{K}}}_i^m(\bm{\theta}_t^m)$;\\
 ${HTA}_t^m= HTA\left({{{K}}}^m(\bm{\theta}_t^m),{K}^{*}\middle|\lambda=\lambda_0\right)$;\\
 Compute $\{ \nabla\kappa_{ \bm{\theta}^m} (\bm{x}_i^m,\bm{x}_j^m)\}_{i,j=1}^{N}$ on the quantum computing device;\\

 Compute the gradient $g_t=g_{HTA}(\bm{\theta}^m_{t})$ according to Eq.~(\ref{eq:gta}) and (\ref{eq:ghta});\\
 $\bm{\theta}_{t+1}^m =\bm{\theta}_t^m+\xi g_t$;\\
 $t = t+1$;\\}
 
 }
 \textbf{Return: } ${{\{{{K}}}^m}\}_{m=1}^M$, $\hat{\bm{\theta}}=\{ \hat{\bm{\theta}}^1,\cdots,\hat{\bm{\theta}}^M \}$. \\
 
 \setcounter{AlgoLine}{0}
  \nonl \textbf{Stage 2:} Training weight coefficients of the quantum multi-kernel \\
 \KwIn { ${{\{{{K}}}^m}\}_{m=1}^M$, $\bm{\eta}_{\text{random}}$, $T_2$, $\lambda_0$, $k_2$, and $\epsilon_2$.} 
 \KwOut { $\bm{\eta}^{*}$. }
 Initialize $\bm{\eta}_1=\bm{\eta}_{random}$, ${HTA}_0^C=0$, and $t=1$;\\
 \Repeat{$\left\|{HTA}^C_t-{HTA}^C_{t-1} \right\|\le\epsilon_2$ or $t=T_2$}{
 Compute $\sum_{m=1}^M \eta^m_t K^m$, and obtain the $k_2$ nearest neighbors $\mathcal{N}^c_{i(t)}$ for each instance;\\
 According to $\mathcal{N}^c (\bm{x}_i)$ and corresponding class labels, construct $K_i^{*}$;\\
 ${HTA}_t^C = HTA (K^c_{\bm{\eta}_t,\hat{\bm{\theta}}},K^{*} |\lambda=\lambda_0 )$;\\
 Fixing $\bm{\tau}$, update $\bm{\mu}$ according to Eq.~(\ref{eq:2ob});\\
 $\eta^m_t = \mu^m_t/\sum_m \mu^m_t$;\\
 Fixing $\bm{\eta}_t$, compute $\bm{\tau}$; \\}
 \textbf{Return: } $\bm{\eta}^{*}$.
\end{algorithm}

\section{Numerical simulations}
\label{sec:4}
\par We evaluate the proposed quantum multi-view kernel learning model on the Mfeat dataset~\cite{mfeat_dataset} through numerical simulations implemented via the PennyLane~\cite{bergholm2018pennylane} quantum simulation platform. 
\par For implementation efficiency, during the training process, the base kernel parameters are optimized using Stochastic Gradient Descent (SGD)~\cite{amari1993backpropagation}, and the cost functions share the same hyperparameter with that of the quantum multi-kernel optimization. The resulting optimized combined kernel is then applied to the SVM classifier, a canonical kernel-based approach, and we assess its accuracy on image recognition tasks. To eliminate randomness, all experiments are repeated 20 times.
\par Under the aforementioned configuration, we conduct experiments to assess the performance of our algorithm from various perspectives, including performance under different hybrid parameters $\lambda$ and numbers of neighbors $k$, performance under different quantum circuit depths $P$, and comparative performance against classical counterparts. 

\subsection{Dataset}
\par Mfeat dataset is a representative multi-feature dataset of handwritten numerals (`0' - `9'), comprising 6 pre-extracted features. This dataset contains 200 instances per class, totaling 2,000 instances. In this paper, we conduct experiments to classify digits into two categories: `0' - `4' ($-1$) and `5' - `9' ($+1$). Furthermore, we use Principal Component Analysis (PCA) to reduce the dimension of each feature to $6$. Detailed information about the Mfeat dataset is shown in TABLE~\ref{tab:dataset}. Besides, we randomly select 80 samples for the training set and 80 samples for the test set, with equal representation from both classes (40 samples per class in each subset) to avoid class imbalance bias.

\begin{table}[ht]
\footnotesize
 \centering
 \caption{Features of the Mfeat dataset. $d$ and $d_r$ denote the original dimension and the reduced dimension of each feature, respectively.}
 \begin{tabular}{ccccl}
 \hline 
 Feature & $d$ & $d_r$ & \multicolumn{1}{c}{Describe}\\
 \hline
 FOU & 76 & 6& Fourier coefficients of the character shapes\\
 FAC& 216 & 6&Profile correlations\\
 KAR& 64 & 6&Karhunen-Love coefficients\\
 PIX & 240& 6&Pixel averages in 2 $\times$ 3 windows\\
 ZER& 47& 6&Zernike moments\\
 MOR& 6& 6&Morphological features\\
 \hline
 \end{tabular}
 \label{tab:dataset}
\end{table}

\subsection{Structure of quantum base kernel}

\par
\begin{figure}[ht]
 \centering
 \includegraphics[width = 8.2cm]{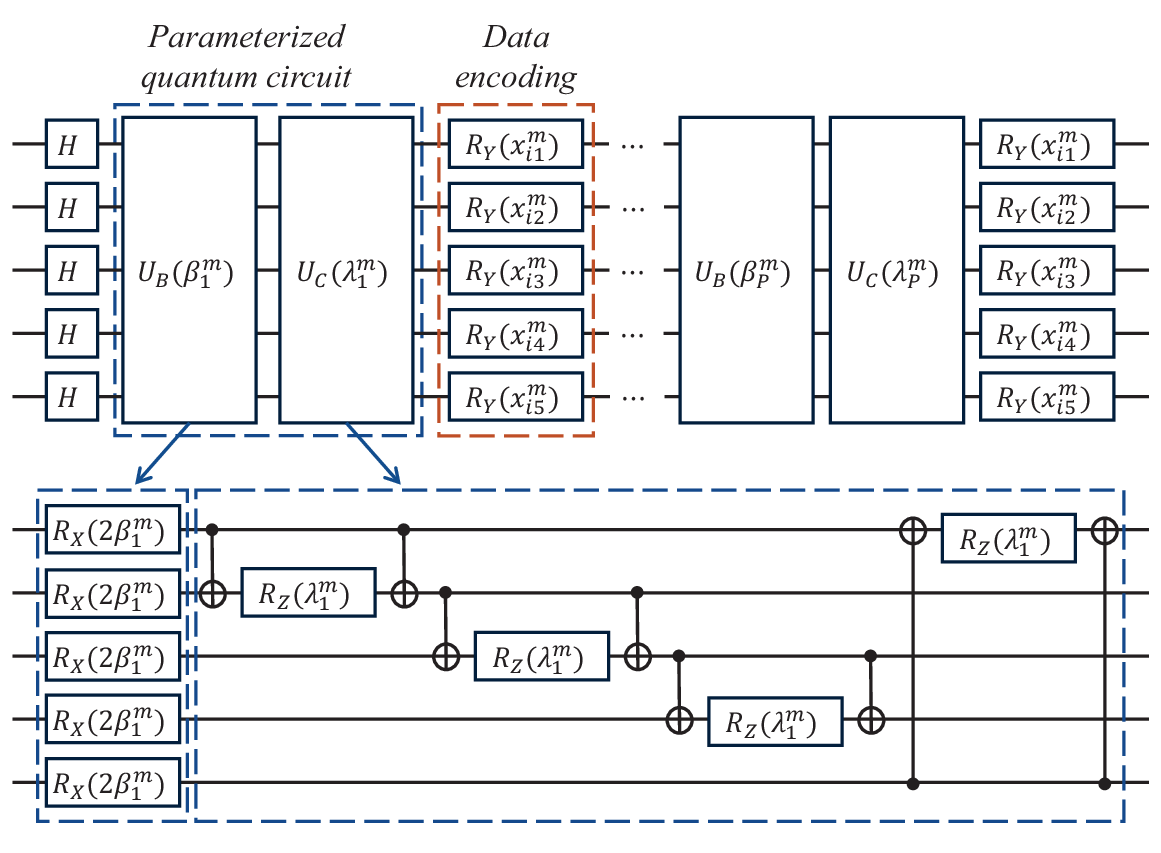}
 \caption{Quantum circuit of $W (\bm{x}_i^m,\bm{\theta}^m)$.}
 \label{fig:pqc}
\end{figure}

 For ease of implementation, in our experiments, the quantum base kernels adopt the same construction by the operation $W (\bm{x}_i^m,\bm{\theta}^m)$. As shown in FIG.~\ref{fig:pqc}, $W (\bm{x}_i^m,\bm{\theta}^m)$ consists of a single-layer Hadamard gates $H$ and $P$-layer parameterized data-encoding operations. Each parameterized data-encoding operation is composed of a PQC and a data-encoding block. 

\par For the $m$-th feature vector $\bm{x}_i^m = (x_{i1}^m,\cdots,x_{id_m}^m)$, data-encoding block embeds the classical vector into quantum states by the rotation Pauli gates $R_Y (x_{iq}^m)$. Denoting the data-encoding block as $D (\bm{x}_i^m)$, we have
\begin{equation}
 D (\bm{x}_i^m) = \prod_{q=1}^{d_m} R_Y (x_{iq}^m)_q.
\end{equation}

Based on QAOA, a widely used circuit structure in QML with fewer parameters and high expressiveness, PQC is composed of two unitaries,
\begin{align}
U_B (\beta_p^m) &= e^{-i \beta_p^m H_B}, \\
U_C (\gamma_p^m) &= e^{-i \gamma_p^m H_C},
\end{align}
where the Hamiltonians $H_B= \sum_{q=1}^{d_m} \sigma_q^x$ and $H_C= \sum_{q=1}^{d_m} \frac{1}{2} (I-\sigma_q^z \sigma_{q+1}^z)$. $\sigma_q^x$ and $\sigma_q^x$ represent the Pauli $X$ operator and the Pauli $Z$ operator acting on the $q$-th qubit respectively. $ U_B (\beta_p^m)$ and $U_C (\gamma_p^m)$ can be implemented by the rotation Pauli gates $R_X (2\beta_p^m)$ and $R_Z (\gamma_p^m)$,
\begin{gather}
 U_B (\beta_p^m) = \prod_{q=1}^{d_m} R_X (2\beta_p^m)_q, \\
U_C (\gamma_p^m) = \prod_{q=1}^{d_m} CNOT_{ (q,q+1)} R_Z (\gamma_p^m)_{q+1} CNOT_{ (q,q+1)}.
\end{gather} 
Overall, 
\begin{equation}
 W (\bm{x}_i^m,\bm{\theta}^m) = H^{\otimes {d_m}} \prod_{p=1}^P ( U_B (\beta_p^m) U_C ( \gamma_p^m) D (\bm{x}_i^m)),
\end{equation}
where $\bm{\theta}^m = (\beta_1^m,\cdots, \beta_P^m, \gamma_1^m,\cdots, \gamma_P^m)$ is the parameters of the base kernel.

\par
\begin{figure}[ht]
 \centering
 \includegraphics[width= 5.5 cm]{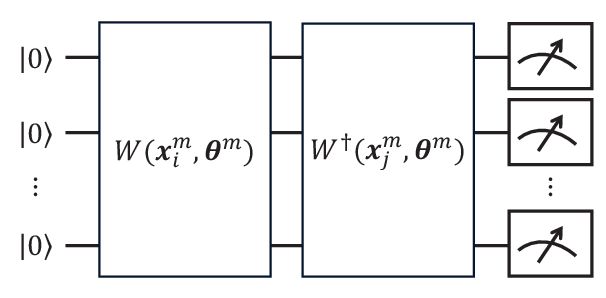}
 \caption{Adjoint approach for estimating $\kappa_{\bm{\theta}^m}^m (\bm{x}_i^m,\bm{x}_j^m)$.}
 \label{fig:inner}
\end{figure}

 After constructing $W (\bm{x}_i^m,\bm{\theta}^m)$, the value of the quantum kernel is estimated by the adjoint approach as shown in FIG.~\ref{fig:inner}. Here, to estimate a single quantum kernel $\kappa_{\bm{\theta}^m}^m (\bm{x}_i^m,\bm{x}_j^m)$, $d_m$ qubits, $2d_m (1+3P)$ single-qubit gates and $4 (d_m-1)$ $CNOT$ gates are required in one quantum program execution.

\subsection{Performance under different hybrid parameters and numbers of neighbors}

\begin{figure*}
 \centering
 \includegraphics[width = 13.5 cm]{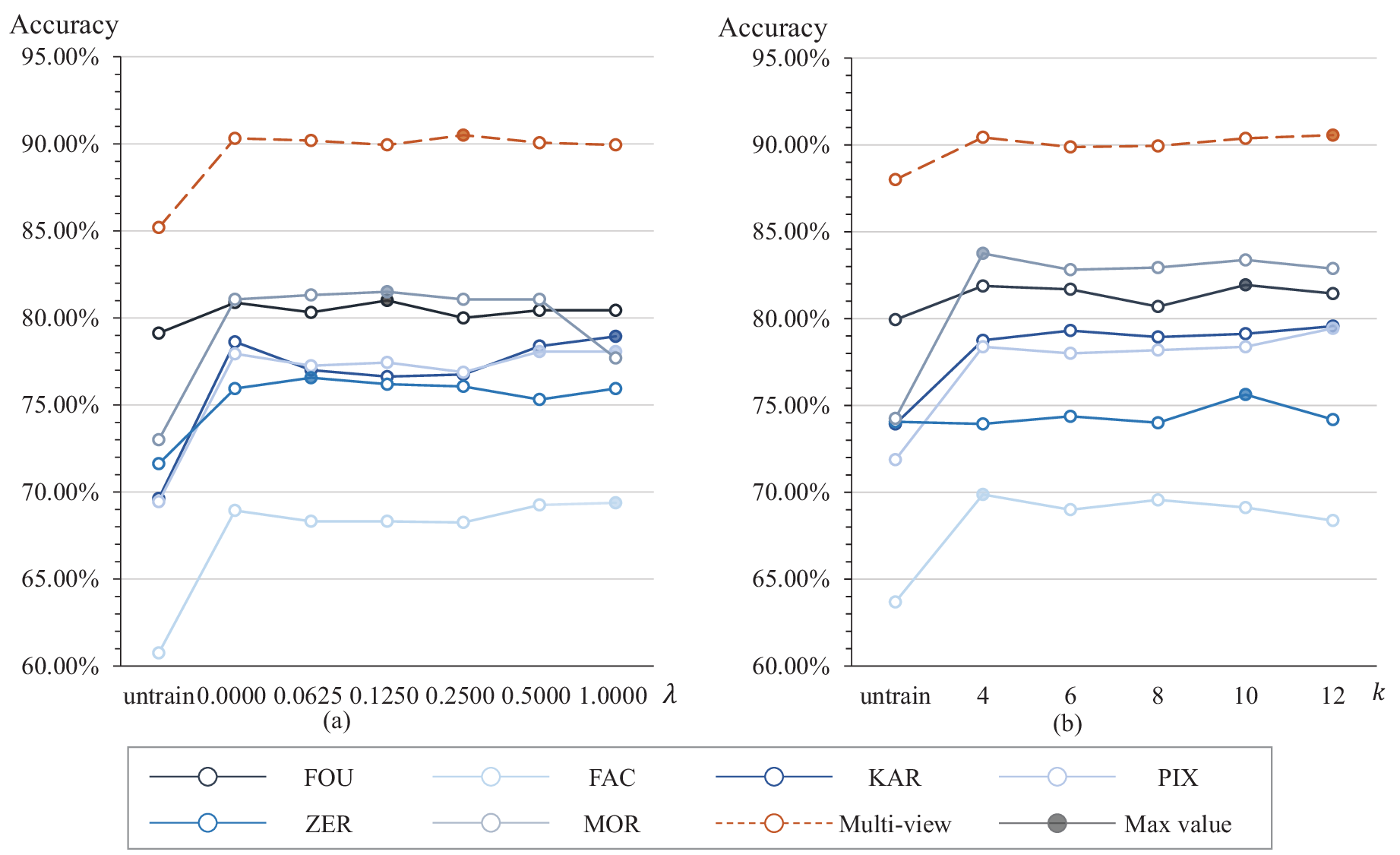}
 \caption{Mean accuracy under different (a) hybrid parameters $\lambda$ and (b) numbers of neighbors $k$.}
 \label{fig:kta}
\end{figure*}

\par In this experiment, we compare the mean accuracy of our multi-view kernel model to that of the single-view kernel model using the same quantum kernel structure. Meanwhile, to evaluate the effects of the local information, we assess the performance under different hybrid parameters $\lambda$ and numbers of neighbors $k$ in hybrid global-local kernel-target alignment. We fix one parameter and observe the other. Besides, we set the quantum depths to be $6$. FIG.~\ref{fig:kta} shows the mean accuracy of the multi-view kernel model and our multi-view kernel model under different $\lambda$ and $k$. The concrete numerical results are shown in TABLE~\ref{tab:a1} and \ref{tab:a2}.
\par According to that, we come to the results as follows.
\begin{itemize}
\item Multi-view models achieve significantly higher accuracy than single-view models. Specifically, at $k=4$, the L-QMVKL model achieves 6.69\% higher accuracy than the peak accuracy of single-view models.

\item The utilization of local information improves the accuracy of the model. When $\lambda=0.25$, the L-QMVKL model achieves 0.19\% and 0.56\% higher accuracy than the cases based on local information only ($\lambda=0$) and global information only ($\lambda=1$), respectively. In addition, the local information has a positive effect on the training of part of the base kernels. In particular, for MOR features, the model accuracy based on hybrid global-local information is 3.81\% higher than the global case, and 6.5\% higher than the case where quantum base kernels are not trained.
\item The performance of L-QMVKL gets improved as the number of nearest neighbor points increases, up to 90.56\% accuracy at $k=12$.

\end{itemize}
 \begin{table*}
\footnotesize
 \caption{Accuracy (mean $\pm$ std (unit: \%)) under different hybrid parameters $\lambda$.}
 \centering
 \begin{tabular}{ccccccccc}
 \toprule
 \multirow{2}{*}{ } & \multirow{2}{*}{Feature} &\multirow{2}{*}{Untrained} & \multicolumn{6}{c}{Hybrid parameter $\lambda$} \\ \cmidrule(l){4-9}
 & & & 0 & 0.0625 & 0.125 & 0.25 & 0.5 & 1 \\
 \midrule
 \multirow{6}{*}{Single-view}& {FOU} & $79.13\pm4.78$&$80.88\pm5.69$&$80.31\pm5.33$&$\bm{81.00}\pm \bm{6.08}$&$80.00\pm6.39$&$80.44\pm5.59$&$80.44\pm4.88$ \\
 &{FAC}& $60.75\pm6.32$&$68.94\pm5.96$&$68.31\pm5.84$&$68.31\pm6.05$&$68.25\pm6.52$&$69.25\pm6.46$&$\bm{69.38}\pm \bm{6.51}$ \\ 
 &{KAR} & $69.63\pm4.26$&$78.63\pm4.89$&$77.00\pm6.69$&$76.63\pm6.82$&$76.75\pm6.10$&$78.38\pm5.77$&$\bm{78.94}\pm \bm{4.76}$ \\
 &{PIX} & $69.44\pm4.99$&$77.94\pm6.10$&$77.25\pm6.61$&$77.44\pm6.12$&$76.88\pm7.55$&$\bm{78.06}\pm \bm{6.07}$&$\bm{78.06}\pm \bm{5.73}$ \\ 
 &{ZER} & $71.63\pm4.28$&$75.94\pm6.36$&$\bm{76.56}\pm \bm{4.07}$&$76.19\pm4.70$&$76.06\pm4.53$&$75.31\pm4.92$&$75.94\pm6.54$ \\ 
 &{MOR} & $73.00\pm5.73$&$81.06\pm5.39$&$81.31\pm5.97$&$\bm{81.50}\pm \bm{6.37}$&$81.06\pm5.97$&$81.06\pm5.59$&$77.69\pm6.15$ \\
 \midrule
 {Multi-view} & {-} & $85.19\pm5.13$&$90.31\pm3.83$&$90.19\pm4.07$&$89.94\pm4.14$&$\bm{90.50}\pm \bm{4.67}$&$90.06\pm4.86$&$89.94\pm3.68$ \\
 \bottomrule
 \end{tabular}
 \label{tab:a1}
\end{table*}

\begin{table*}
\footnotesize
 \caption{Accuracy (mean $\pm$ std (unit: \%)) under different numbers of neighbors $k$.}
 \centering
 \begin{tabular}{cccccccc}
 \toprule
 \multirow{2}{*}{ } & \multirow{2}{*}{Feature} &\multirow{2}{*}{Untrained} & \multicolumn{5}{c}{Number of neighbors $k$} \\ 
 \cmidrule(l){4-8}
 & & & 4 & 6 & 8 & 10 & 12 \\
 \midrule
 \multirow{6}{*}{Single-view}& {FOU} & $79.94\pm 4.62$&$81.88\pm4.25$&$81.69\pm4.68$&$80.69\pm4.58$&$\bm{81.94}\pm \bm{4.91}$&$81.44\pm4.85$ \\ 
 &{FAC} & $63.69\pm6.99$&$\bm{69.88}\pm\bm{6.72}$&$69.00\pm7.24$&$69.56\pm6.90$&$69.13\pm7.05$&$68.38\pm7.43$ \\
 
 &{KAR} & $73.94\pm5.79$&$78.75\pm5.38$&$79.31\pm4.28$&$78.94\pm4.88$&$79.13\pm4.97$&$\bm{79.56}\pm \bm{5.08}$ \\
 
 &{PIX} & $71.88\pm5.16$&$78.38\pm3.92$&$78.00\pm4.00$&$78.19\pm4.15$&$78.38\pm4.51$&$\bm{79.44}\pm \bm{4.77}$ \\
 
 &{ZER} & $74.06\pm4.16$&$73.94\pm6.43$&$74.38\pm6.08$&$74.00\pm6.54$&$\bm{75.63}\pm \bm{4.62}$&$74.19\pm5.99$ \\
 
 &{MOR} & $74.25\pm5.45$&$\bm{83.75}\pm\bm{3.47}$&$82.81\pm3.55$&$82.94\pm3.69$&$83.38\pm 3.49$&$82.88\pm4.51$ \\
 
 \midrule
 {Multi-view} & {-} & $88.00\pm4.75$&$90.44\pm4.23$&$89.88\pm4.29$&$89.94\pm4.19$&$90.38\pm4.17$& {$\bm{90.56}\pm \bm{3.82}$}\\
 \bottomrule
 \end{tabular}
 \label{tab:a2}
\end{table*}

\par Through leveraging the local information to train the quantum base kernel, and effectively combining the optimized view-specific quantum kernels, L-QMVKL exhibits a promising performance. In our numerical simulations, each base kernel is trained with the same hyperparameters. According to experimental results, we observe that the optimal hyperparameters corresponding to the maximum accuracy vary across different feature views. Different views exhibit distinct sensitivities to these parameters. Therefore, in practical applications, the algorithm performance can be further improved by adaptive hyperparameter selection tailored to specific feature views.

\subsection{Performance under different quantum circuit depths}
\par
\begin{figure}[ht]
 \centering
 \includegraphics[scale = 0.3]{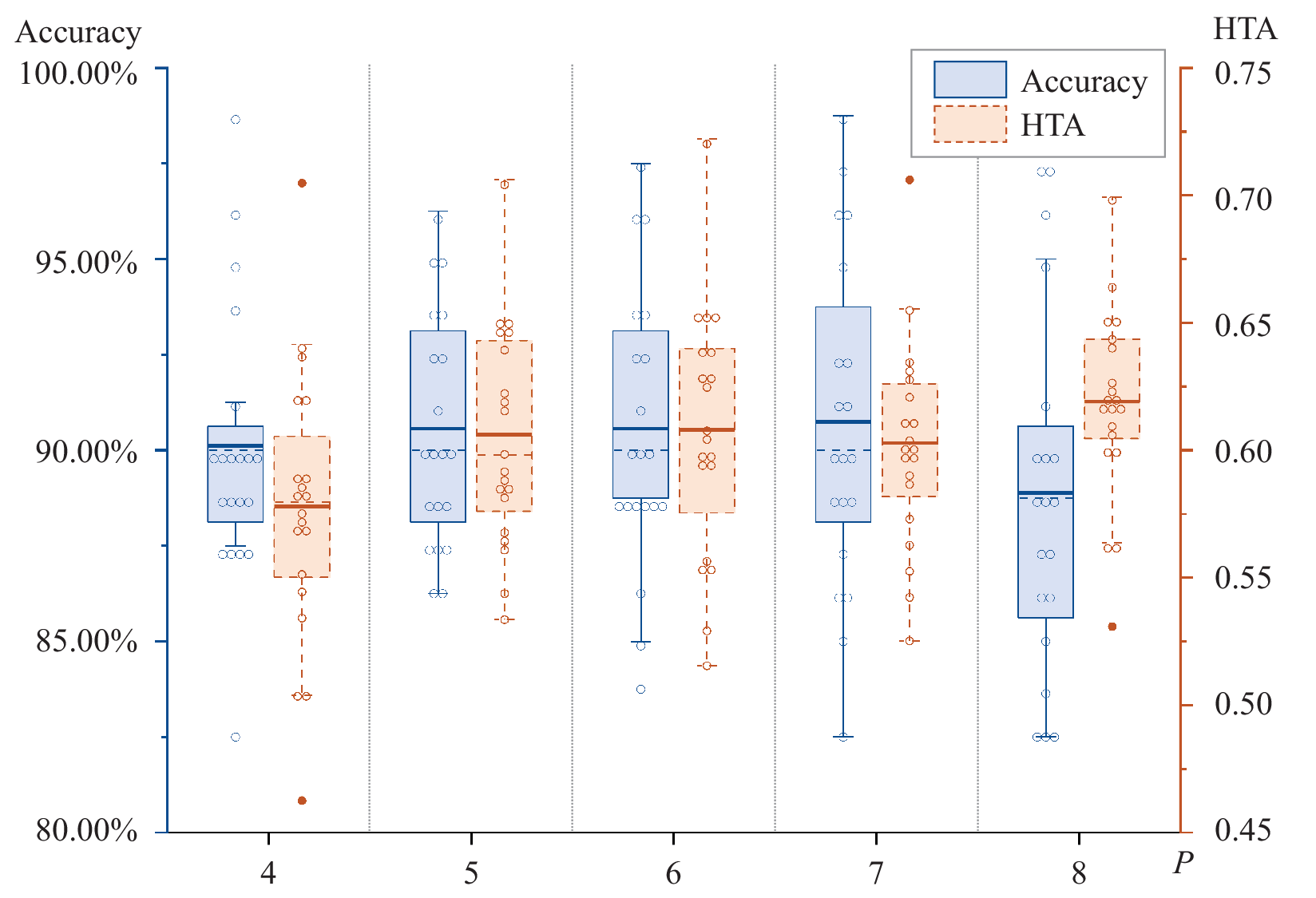}
 \caption{Performance under different circuit depth $P$.}
 \label{fig:depth}
\end{figure}

 The depth of the quantum circuit (i.e., the number of layers) is a factor that affects algorithm performance. To analyze its impact, we execute experiments with varying depths, ranging from $4$ to $8$ layers. Here, $\lambda$ and $k$ are fixed to $0.125$ and $8$, respectively. FIG.~\ref{fig:depth} and TABLE~\ref{tab:a3} show the act under different $P$. 

\par According to that, we come to the results as follows.
\begin{itemize}
 \item For the accuracy, as the depth $P$ increases, there is an overall upward trend. Particularly, when $P=7$, the mean accuracy is 90.75\%. However, when $P=8$, the average accuracy drops to 88.75\%.
 \item The value of $HTA (K^c, K^{*})$ increases with $P$ growing.
\end{itemize} 

\begin{table*}
\footnotesize
 \caption{Accuracy (mean $\pm$ std (unit: \%)) under different circuit depth $P$.}
 \centering
 \begin{tabular}{ccccccc}
 \toprule
 \multirow{2}{*}{ } & \multirow{2}{*}{Feature} & \multicolumn{5}{c}{Circuit depth $P$} \\ 
 \cmidrule(l){3-7}
 &  & 4 & 5 & 6 & 7 & 8 \\
 \midrule
 \multirow{6}{*}{Single-view}& {FOU} &$81.88\pm4.77$&$\bm{82.44}\pm\bm{3.78}$&$82.25\pm4.10$&${80.88}\pm {4.78}$&$81.06\pm4.04$ \\ 
 &{FAC} &${64.31}\pm{5.62}$&$66.69\pm6.93$&$65.31\pm7.39$&$66.25\pm7.05$&$\bm{68.56}\pm\bm{6.79}$ \\
 
 &{KAR} &$77.63\pm6.70$&$\bm{79.69}\pm\bm{5.97}$&$79.06\pm6.38$&$79.38\pm5.05$&$\bm{79.69}\pm \bm{6.25}$ \\
 
 &{PIX} &$76.25\pm5.53$&$\bm{78.81}\pm\bm{5.44}$&$77.75\pm6.37$&$76.94\pm5.91$&${78.25}\pm {6.68}$ \\
 
 &{ZER} &$76.88\pm5.19$&$76.94\pm4.93$&$76.38\pm4.81$&${76.38}\pm {5.30}$&$\bm{78.56}\pm \bm{4.35}$ \\
 
 &{MOR} &${79.38}\pm{4.72}$&$80.81\pm4.93$&$81.31\pm5.60$&$81.56\pm 4.29$&$\bm{82.69}\pm\bm{5.07}$ \\
 
 \midrule
 {Multi-view} & {-} &$90.13\pm3.49$&$90.56\pm2.97$&$80.56.94\pm3.59$&$\bm{90.75}\pm\bm{4.25}$& {${88.88}\pm {4.64}$}\\
 \bottomrule
 \end{tabular}
 \label{tab:a3}
\end{table*}

\subsection{Comparative performance against classical counterparts}
\par
\begin{figure}[ht]
 \centering
 \includegraphics[scale = 0.3]{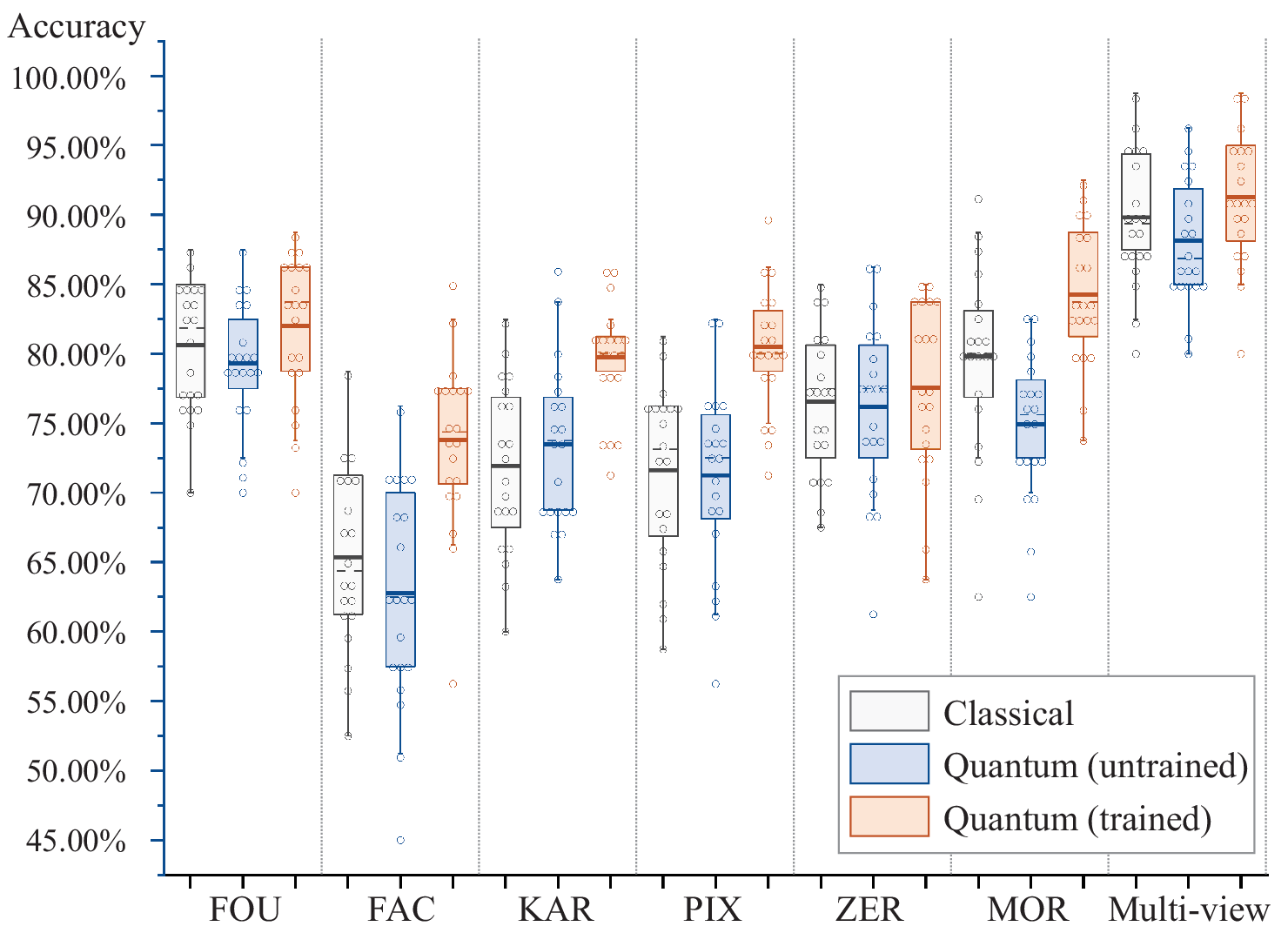}
 \caption{Accuracy of L-QMVKL compared to its classical counterparts.}
 \label{fig:acc}
\end{figure}

 We further compare our quantum model to its classical counterparts. In the classical model, we use the Gaussian kernel as the base kernel, where the hyperparameter is the mean Euclidean distance between sample vectors of each view. In quantum models, $\lambda$, $k$ and $P$ are fixed to $0.125$, $8$ and $6$, respectively. The results are shown in FIG.~\ref{fig:acc} and TABLE~\ref{acc_mfeat}. 

\par According to that, we come to the results as follows. 
\begin{itemize}
\item In the classical scenario, the multi-view learning model also outperforms single-view models.
\item Although untrained quantum base kernels do not consistently surpass classical counterparts in performance, all trained quantum base kernels demonstrate significant superiority over classical kernels.
\item L-QMVKL outperforms its classical counterparts, achieving a 1.44\% higher average accuracy.
\end{itemize}

\begin{table*}
\footnotesize
\centering
\caption{Accuracy and $HTA(\cdot, K^{*})$ (mean $\pm$ std (unit: \%)) compared to its classical counterparts.}
\label{acc_mfeat}
\begin{tabular}{ccccc}
\toprule
 &\multirow{2}{*}{Feature} & \multirow{2}{*}{Classical} & \multicolumn{2}{c}{Quantum kernel} \\
 \cmidrule(l){4-5}
 & & & Untrained
 & Trained \\
\midrule
\multirow{12}{*}{Single-view} & \multirow{2}{*}{FOU} & 80.63 $\pm$ 4.57 &79.31 $\pm$ 4.45 &\textbf{82.00} $\pm$ \textbf{5.10} \\
 & &(28.20 $\pm$ 10.50) &(28.30 $\pm$ 10.65) & (36.47 $\pm$ 9.68) \\
 & \multirow{2}{*}{FAC} &65.38 $\pm$ 6.33 & 62.75 $\pm$ 7.73 &\textbf{73.81} $\pm$ \textbf{6.10} \\
 & &(12.87 $\pm$ 5.45) &(12.19 $\pm$ 5.35) &(21.45 $\pm$ 5.92) \\
 & \multirow{2}{*}{KAR} &71.94 $\pm$ 5.92 &73.50 $\pm$ 5.74 &\textbf{79.75} $\pm$ \textbf{3.96} \\
 & & (21.12 $\pm$ 7.54) &(21.52 $\pm$ 7.07) &(33.00 $\pm$ 6.59) \\
 & \multirow{2}{*}{PIX} &71.63 $\pm$ 6.26 & 71.25 $\pm$ 6.52 &\textbf{80.50} $\pm$ \textbf{4.39} \\
 & &(22.14 $\pm$ 7.05) &(22.47 $\pm$ 6.62) &(33.93 $\pm$ 5.80) \\
 & \multirow{2}{*}{ZER} &76.56 $\pm$ 4.94 & 76.19 $\pm$ 6.14 &\textbf{77.56} $\pm$ \textbf{6.08} \\
 & & (20.42 $\pm$ 7.18) &(19.84 $\pm$ 7.12) &(29.67 $\pm$ 6.55) \\
 & \multirow{2}{*}{MOR} &79.81 $\pm$ 6.50 & 74.94 $\pm$ 5.10 & \textbf{84.25} $\pm$ \textbf{4.86} \\
 & &(39.98 $\pm$ 7.84) &(39.06 $\pm$ 7.05) & (42.90 $\pm$ 6.82) \\
\midrule
\multirow{2}{*}{Multi-view} 
 & \multirow{2}{*}{-} & 89.81 $\pm$ 4.63 & 88.13 $\pm$ 4.38 &\textbf{91.25 $\pm$ 4.56} \\
 & &(58.11 $\pm$ 3.53) &(53.16 $\pm$ 6.32) &(62.10 $\pm$ 4.55) \\
\bottomrule
\end{tabular}
\end{table*}
\section{Conclusions}
\label{sec:6}
\par In this paper, we propose quantum multi-view kernel learning with local information, L-QMVKL. L-QMVKL combines optimized view-specific quantum kernels to construct the quantum multi-kernel, characterizing and fusing the features across diverse views of data. With the optimization goal of maximizing the hybrid global-local kernel alignment, the parameters of quantum base kernels and weight coefficients in L-QMVKL are trained, enabling the effective capture of the local structure. We validate the performance of L-QMVKL through comprehensive numerical simulations. Results show that L-QMVKL achieves a superior accuracy over single-view models, and the utilization of local information provides a positive effect for parameter training. Besides, our algorithm also outperforms its classical counterpart. 

\par Our work enhances the ability of quantum kernel methods to handle complex data, and is expected to drive deeper research in the field of quantum machine learning. Two critical challenges demand urgent attention in advancing kernel-based quantum algorithms. The first challenge lies in the high computational complexity of kernel matrix construction, which remains a fundamental bottleneck for scaling kernel methods. Developing advanced techniques to accelerate this process is essential for improving the practicality of kernel-based algorithms.
The second challenge stems from the quantum noise. Quantum noise mitigation is a cornerstone of reliable quantum computing~\cite{preskill2018quantum,cai2023quantum}. By systematically combining it with the quantum kernels, it helps to improve the robustness in practical applications. 
Addressing these challenges will be pivotal in unlocking the full potential of quantum kernel methods for practical machine learning tasks. In the future, we will conduct in-depth investigations around these aspects. 

\section*{Data Availability Statement}
The data are available upon reasonable request from the authors.

\section*{Acknowledgments}
This work is supported by National Natural Science Foundation of China (Grant Nos. 62372048, 62371069, 62272056) and Tianyan Quantum Computing Program. 


\bibliography{main.bib}

\begin{thebibliography}{57}%
\makeatletter
\providecommand \@ifxundefined [1]{%
 \@ifx{#1\undefined}
}%
\providecommand \@ifnum [1]{%
 \ifnum #1\expandafter \@firstoftwo
 \else \expandafter \@secondoftwo
 \fi
}%
\providecommand \@ifx [1]{%
 \ifx #1\expandafter \@firstoftwo
 \else \expandafter \@secondoftwo
 \fi
}%
\providecommand \natexlab [1]{#1}%
\providecommand \enquote  [1]{``#1''}%
\providecommand \bibnamefont  [1]{#1}%
\providecommand \bibfnamefont [1]{#1}%
\providecommand \citenamefont [1]{#1}%
\providecommand \href@noop [0]{\@secondoftwo}%
\providecommand \href [0]{\begingroup \@sanitize@url \@href}%
\providecommand \@href[1]{\@@startlink{#1}\@@href}%
\providecommand \@@href[1]{\endgroup#1\@@endlink}%
\providecommand \@sanitize@url [0]{\catcode `\\12\catcode `\$12\catcode `\&12\catcode `\#12\catcode `\^12\catcode `\_12\catcode `\%12\relax}%
\providecommand \@@startlink[1]{}%
\providecommand \@@endlink[0]{}%
\providecommand \url  [0]{\begingroup\@sanitize@url \@url }%
\providecommand \@url [1]{\endgroup\@href {#1}{\urlprefix }}%
\providecommand \urlprefix  [0]{URL }%
\providecommand \Eprint [0]{\href }%
\providecommand \doibase [0]{https://doi.org/}%
\providecommand \selectlanguage [0]{\@gobble}%
\providecommand \bibinfo  [0]{\@secondoftwo}%
\providecommand \bibfield  [0]{\@secondoftwo}%
\providecommand \translation [1]{[#1]}%
\providecommand \BibitemOpen [0]{}%
\providecommand \bibitemStop [0]{}%
\providecommand \bibitemNoStop [0]{.\EOS\space}%
\providecommand \EOS [0]{\spacefactor3000\relax}%
\providecommand \BibitemShut  [1]{\csname bibitem#1\endcsname}%
\let\auto@bib@innerbib\@empty
\bibitem [{\citenamefont {LeCun}\ \emph {et~al.}(2015)\citenamefont {LeCun}, \citenamefont {Bengio},\ and\ \citenamefont {Hinton}}]{lecun2015deep}%
  \BibitemOpen
  \bibfield  {author} {\bibinfo {author} {\bibfnamefont {Y.}~\bibnamefont {LeCun}}, \bibinfo {author} {\bibfnamefont {Y.}~\bibnamefont {Bengio}},\ and\ \bibinfo {author} {\bibfnamefont {G.}~\bibnamefont {Hinton}},\ }\bibfield  {title} {\bibinfo {title} {Deep learning},\ }\href {https://doi.org/10.1038/nature14539} {\bibfield  {journal} {\bibinfo  {journal} {Nature}\ }\textbf {\bibinfo {volume} {521}},\ \bibinfo {pages} {436} (\bibinfo {year} {2015})}\BibitemShut {NoStop}%
\bibitem [{\citenamefont {Jordan}\ and\ \citenamefont {Mitchell}(2015)}]{jordan2015machine}%
  \BibitemOpen
  \bibfield  {author} {\bibinfo {author} {\bibfnamefont {M.~I.}\ \bibnamefont {Jordan}}\ and\ \bibinfo {author} {\bibfnamefont {T.~M.}\ \bibnamefont {Mitchell}},\ }\bibfield  {title} {\bibinfo {title} {Machine learning: Trends, perspectives, and prospects},\ }\href {https://doi.org/10.1126/science.aaa8415} {\bibfield  {journal} {\bibinfo  {journal} {Science}\ }\textbf {\bibinfo {volume} {349}},\ \bibinfo {pages} {255} (\bibinfo {year} {2015})}\BibitemShut {NoStop}%
\bibitem [{\citenamefont {Hofmann}\ \emph {et~al.}(2008)\citenamefont {Hofmann}, \citenamefont {Sch{\"o}lkopf},\ and\ \citenamefont {Smola}}]{hofmann2008kernel}%
  \BibitemOpen
  \bibfield  {author} {\bibinfo {author} {\bibfnamefont {T.}~\bibnamefont {Hofmann}}, \bibinfo {author} {\bibfnamefont {B.}~\bibnamefont {Sch{\"o}lkopf}},\ and\ \bibinfo {author} {\bibfnamefont {A.}~\bibnamefont {Smola}},\ }\bibfield  {title} {\bibinfo {title} {Kernel methods in machine learning},\ }\href {https://doi.org/10.1214/009053607000000677} {\bibfield  {journal} {\bibinfo  {journal} {Ann. Stat.}\ }\textbf {\bibinfo {volume} {36}},\ \bibinfo {pages} {1171} (\bibinfo {year} {2008})}\BibitemShut {NoStop}%
\bibitem [{\citenamefont {G{\"o}nen}\ and\ \citenamefont {Alpayd{\i}n}(2011)}]{gonen2011multiple}%
  \BibitemOpen
  \bibfield  {author} {\bibinfo {author} {\bibfnamefont {M.}~\bibnamefont {G{\"o}nen}}\ and\ \bibinfo {author} {\bibfnamefont {E.}~\bibnamefont {Alpayd{\i}n}},\ }\bibfield  {title} {\bibinfo {title} {Multiple kernel learning algorithms},\ }\href@noop {} {\bibfield  {journal} {\bibinfo  {journal} {J. Mach. Learn. Res.}\ }\textbf {\bibinfo {volume} {12}},\ \bibinfo {pages} {2211} (\bibinfo {year} {2011})}\BibitemShut {NoStop}%
\bibitem [{\citenamefont {Shor}(1994)}]{shor1994algorithms}%
  \BibitemOpen
  \bibfield  {author} {\bibinfo {author} {\bibfnamefont {P.}~\bibnamefont {Shor}},\ }\bibfield  {title} {\bibinfo {title} {Algorithms for quantum computation: discrete logarithms and factoring},\ }in\ \href {https://doi.org/10.1109/SFCS.1994.365700} {\emph {\bibinfo {booktitle} {Proceedings 35th Annual Symposium on Foundations of Computer Science}}}\ (\bibinfo {year} {1994})\ pp.\ \bibinfo {pages} {124--134}\BibitemShut {NoStop}%
\bibitem [{\citenamefont {Grover}(1997)}]{grover1997quantum}%
  \BibitemOpen
  \bibfield  {author} {\bibinfo {author} {\bibfnamefont {L.~K.}\ \bibnamefont {Grover}},\ }\bibfield  {title} {\bibinfo {title} {Quantum mechanics helps in searching for a needle in a haystack},\ }\href {https://doi.org/10.1103/PhysRevLett.79.325} {\bibfield  {journal} {\bibinfo  {journal} {Phys. Rev. Lett.}\ }\textbf {\bibinfo {volume} {79}},\ \bibinfo {pages} {325} (\bibinfo {year} {1997})}\BibitemShut {NoStop}%
\bibitem [{\citenamefont {Harrow}\ \emph {et~al.}(2009)\citenamefont {Harrow}, \citenamefont {Hassidim},\ and\ \citenamefont {Lloyd}}]{harrow2009quantum}%
  \BibitemOpen
  \bibfield  {author} {\bibinfo {author} {\bibfnamefont {A.~W.}\ \bibnamefont {Harrow}}, \bibinfo {author} {\bibfnamefont {A.}~\bibnamefont {Hassidim}},\ and\ \bibinfo {author} {\bibfnamefont {S.}~\bibnamefont {Lloyd}},\ }\bibfield  {title} {\bibinfo {title} {Quantum algorithm for linear systems of equations},\ }\href {https://doi.org/10.1103/PhysRevLett.103.150502} {\bibfield  {journal} {\bibinfo  {journal} {Phys. Rev. Lett.}\ }\textbf {\bibinfo {volume} {103}},\ \bibinfo {pages} {150502} (\bibinfo {year} {2009})}\BibitemShut {NoStop}%
\bibitem [{\citenamefont {Biamonte}\ \emph {et~al.}(2017)\citenamefont {Biamonte}, \citenamefont {Wittek}, \citenamefont {Pancotti}, \citenamefont {Rebentrost}, \citenamefont {Wiebe},\ and\ \citenamefont {Lloyd}}]{biamonte2017quantum}%
  \BibitemOpen
  \bibfield  {author} {\bibinfo {author} {\bibfnamefont {J.}~\bibnamefont {Biamonte}}, \bibinfo {author} {\bibfnamefont {P.}~\bibnamefont {Wittek}}, \bibinfo {author} {\bibfnamefont {N.}~\bibnamefont {Pancotti}}, \bibinfo {author} {\bibfnamefont {P.}~\bibnamefont {Rebentrost}}, \bibinfo {author} {\bibfnamefont {N.}~\bibnamefont {Wiebe}},\ and\ \bibinfo {author} {\bibfnamefont {S.}~\bibnamefont {Lloyd}},\ }\bibfield  {title} {\bibinfo {title} {Quantum machine learning},\ }\href {https://doi.org/10.1038/nature23474} {\bibfield  {journal} {\bibinfo  {journal} {Nature}\ }\textbf {\bibinfo {volume} {549}},\ \bibinfo {pages} {195} (\bibinfo {year} {2017})}\BibitemShut {NoStop}%
\bibitem [{\citenamefont {Dunjko}\ and\ \citenamefont {Briegel}(2018)}]{dunjko2018machine}%
  \BibitemOpen
  \bibfield  {author} {\bibinfo {author} {\bibfnamefont {V.}~\bibnamefont {Dunjko}}\ and\ \bibinfo {author} {\bibfnamefont {H.~J.}\ \bibnamefont {Briegel}},\ }\bibfield  {title} {\bibinfo {title} {Machine learning \& artificial intelligence in the quantum domain: a review of recent progress},\ }\href {https://doi.org/10.1088/1361-6633/aab406} {\bibfield  {journal} {\bibinfo  {journal} {Rep. Prog. Phys.}\ }\textbf {\bibinfo {volume} {81}},\ \bibinfo {pages} {074001} (\bibinfo {year} {2018})}\BibitemShut {NoStop}%
\bibitem [{\citenamefont {Houssein}\ \emph {et~al.}(2022)\citenamefont {Houssein}, \citenamefont {Abohashima}, \citenamefont {Elhoseny},\ and\ \citenamefont {Mohamed}}]{houssein2022machine}%
  \BibitemOpen
  \bibfield  {author} {\bibinfo {author} {\bibfnamefont {E.~H.}\ \bibnamefont {Houssein}}, \bibinfo {author} {\bibfnamefont {Z.}~\bibnamefont {Abohashima}}, \bibinfo {author} {\bibfnamefont {M.}~\bibnamefont {Elhoseny}},\ and\ \bibinfo {author} {\bibfnamefont {W.~M.}\ \bibnamefont {Mohamed}},\ }\bibfield  {title} {\bibinfo {title} {Machine learning in the quantum realm: The state-of-the-art, challenges, and future vision},\ }\href {https://doi.org/10.1016/j.eswa.2022.116512} {\bibfield  {journal} {\bibinfo  {journal} {Expert Syst. Appl.}\ }\textbf {\bibinfo {volume} {194}},\ \bibinfo {pages} {116512} (\bibinfo {year} {2022})}\BibitemShut {NoStop}%
\bibitem [{\citenamefont {Huang}\ \emph {et~al.}(2023)\citenamefont {Huang}, \citenamefont {Xu}, \citenamefont {Guo}, \citenamefont {Tian}, \citenamefont {Wei}, \citenamefont {Sun}, \citenamefont {Bao},\ and\ \citenamefont {Long}}]{huang2023near}%
  \BibitemOpen
  \bibfield  {author} {\bibinfo {author} {\bibfnamefont {H.-L.}\ \bibnamefont {Huang}}, \bibinfo {author} {\bibfnamefont {X.-Y.}\ \bibnamefont {Xu}}, \bibinfo {author} {\bibfnamefont {C.}~\bibnamefont {Guo}}, \bibinfo {author} {\bibfnamefont {G.}~\bibnamefont {Tian}}, \bibinfo {author} {\bibfnamefont {S.-J.}\ \bibnamefont {Wei}}, \bibinfo {author} {\bibfnamefont {X.}~\bibnamefont {Sun}}, \bibinfo {author} {\bibfnamefont {W.-S.}\ \bibnamefont {Bao}},\ and\ \bibinfo {author} {\bibfnamefont {G.-L.}\ \bibnamefont {Long}},\ }\bibfield  {title} {\bibinfo {title} {Near-term quantum computing techniques: Variational quantum algorithms, error mitigation, circuit compilation, benchmarking and classical simulation},\ }\href {https://doi.org/10.1007/s11433-022-2057-y} {\bibfield  {journal} {\bibinfo  {journal} {Sci. China-Phys. Mech. Astron.}\ }\textbf {\bibinfo {volume} {66}},\ \bibinfo {pages} {250302} (\bibinfo {year} {2023})}\BibitemShut {NoStop}%
\bibitem [{\citenamefont {Rebentrost}\ \emph {et~al.}(2014)\citenamefont {Rebentrost}, \citenamefont {Mohseni},\ and\ \citenamefont {Lloyd}}]{rebentrost2014quantum}%
  \BibitemOpen
  \bibfield  {author} {\bibinfo {author} {\bibfnamefont {P.}~\bibnamefont {Rebentrost}}, \bibinfo {author} {\bibfnamefont {M.}~\bibnamefont {Mohseni}},\ and\ \bibinfo {author} {\bibfnamefont {S.}~\bibnamefont {Lloyd}},\ }\bibfield  {title} {\bibinfo {title} {Quantum support vector machine for big data classification},\ }\href {https://doi.org/10.1103/PhysRevLett.113.130503} {\bibfield  {journal} {\bibinfo  {journal} {Phys. Rev. Lett.}\ }\textbf {\bibinfo {volume} {113}},\ \bibinfo {pages} {130503} (\bibinfo {year} {2014})}\BibitemShut {NoStop}%
\bibitem [{\citenamefont {Schuld}\ \emph {et~al.}(2020)\citenamefont {Schuld}, \citenamefont {Bocharov}, \citenamefont {Svore},\ and\ \citenamefont {Wiebe}}]{schuld2020circuit}%
  \BibitemOpen
  \bibfield  {author} {\bibinfo {author} {\bibfnamefont {M.}~\bibnamefont {Schuld}}, \bibinfo {author} {\bibfnamefont {A.}~\bibnamefont {Bocharov}}, \bibinfo {author} {\bibfnamefont {K.~M.}\ \bibnamefont {Svore}},\ and\ \bibinfo {author} {\bibfnamefont {N.}~\bibnamefont {Wiebe}},\ }\bibfield  {title} {\bibinfo {title} {Circuit-centric quantum classifiers},\ }\href {https://doi.org/10.1103/PhysRevA.101.032308} {\bibfield  {journal} {\bibinfo  {journal} {Phys. Rev. A}\ }\textbf {\bibinfo {volume} {101}},\ \bibinfo {pages} {032308} (\bibinfo {year} {2020})}\BibitemShut {NoStop}%
\bibitem [{\citenamefont {Abbas}\ \emph {et~al.}(2021)\citenamefont {Abbas}, \citenamefont {Sutter}, \citenamefont {Zoufal}, \citenamefont {Lucchi}, \citenamefont {Figalli},\ and\ \citenamefont {Woerner}}]{abbas2021power}%
  \BibitemOpen
  \bibfield  {author} {\bibinfo {author} {\bibfnamefont {A.}~\bibnamefont {Abbas}}, \bibinfo {author} {\bibfnamefont {D.}~\bibnamefont {Sutter}}, \bibinfo {author} {\bibfnamefont {C.}~\bibnamefont {Zoufal}}, \bibinfo {author} {\bibfnamefont {A.}~\bibnamefont {Lucchi}}, \bibinfo {author} {\bibfnamefont {A.}~\bibnamefont {Figalli}},\ and\ \bibinfo {author} {\bibfnamefont {S.}~\bibnamefont {Woerner}},\ }\bibfield  {title} {\bibinfo {title} {The power of quantum neural networks},\ }\href {https://doi.org/10.1038/s43588-021-00084-1} {\bibfield  {journal} {\bibinfo  {journal} {Nat. Comput. Sci.}\ }\textbf {\bibinfo {volume} {1}},\ \bibinfo {pages} {403} (\bibinfo {year} {2021})}\BibitemShut {NoStop}%
\bibitem [{\citenamefont {Beer}\ \emph {et~al.}(2020)\citenamefont {Beer}, \citenamefont {Bondarenko}, \citenamefont {Farrelly}, \citenamefont {Osborne}, \citenamefont {Salzmann}, \citenamefont {Scheiermann},\ and\ \citenamefont {Wolf}}]{beer2020training}%
  \BibitemOpen
  \bibfield  {author} {\bibinfo {author} {\bibfnamefont {K.}~\bibnamefont {Beer}}, \bibinfo {author} {\bibfnamefont {D.}~\bibnamefont {Bondarenko}}, \bibinfo {author} {\bibfnamefont {T.}~\bibnamefont {Farrelly}}, \bibinfo {author} {\bibfnamefont {T.~J.}\ \bibnamefont {Osborne}}, \bibinfo {author} {\bibfnamefont {R.}~\bibnamefont {Salzmann}}, \bibinfo {author} {\bibfnamefont {D.}~\bibnamefont {Scheiermann}},\ and\ \bibinfo {author} {\bibfnamefont {R.}~\bibnamefont {Wolf}},\ }\bibfield  {title} {\bibinfo {title} {Training deep quantum neural networks},\ }\href {https://doi.org/10.1038/s41467-020-14454-2} {\bibfield  {journal} {\bibinfo  {journal} {Nat. Commun.}\ }\textbf {\bibinfo {volume} {11}},\ \bibinfo {pages} {808} (\bibinfo {year} {2020})}\BibitemShut {NoStop}%
\bibitem [{\citenamefont {Song}\ \emph {et~al.}(2024)\citenamefont {Song}, \citenamefont {Wu}, \citenamefont {Wu}, \citenamefont {Li}, \citenamefont {Wen}, \citenamefont {Qin},\ and\ \citenamefont {Gao}}]{song2024quantum}%
  \BibitemOpen
  \bibfield  {author} {\bibinfo {author} {\bibfnamefont {Y.}~\bibnamefont {Song}}, \bibinfo {author} {\bibfnamefont {Y.}~\bibnamefont {Wu}}, \bibinfo {author} {\bibfnamefont {S.}~\bibnamefont {Wu}}, \bibinfo {author} {\bibfnamefont {D.}~\bibnamefont {Li}}, \bibinfo {author} {\bibfnamefont {Q.}~\bibnamefont {Wen}}, \bibinfo {author} {\bibfnamefont {S.}~\bibnamefont {Qin}},\ and\ \bibinfo {author} {\bibfnamefont {F.}~\bibnamefont {Gao}},\ }\bibfield  {title} {\bibinfo {title} {A quantum federated learning framework for classical clients},\ }\href {https://doi.org/10.1007/s11433-023-2337-2} {\bibfield  {journal} {\bibinfo  {journal} {Sci. China-Phys. Mech. Astron.}\ }\textbf {\bibinfo {volume} {67}},\ \bibinfo {pages} {250311} (\bibinfo {year} {2024})}\BibitemShut {NoStop}%
\bibitem [{\citenamefont {Wu}\ \emph {et~al.}(2025{\natexlab{a}})\citenamefont {Wu}, \citenamefont {Li}, \citenamefont {Song}, \citenamefont {Qin}, \citenamefont {Wen},\ and\ \citenamefont {Gao}}]{wu2025fuzzy}%
  \BibitemOpen
  \bibfield  {author} {\bibinfo {author} {\bibfnamefont {S.}~\bibnamefont {Wu}}, \bibinfo {author} {\bibfnamefont {R.}~\bibnamefont {Li}}, \bibinfo {author} {\bibfnamefont {Y.}~\bibnamefont {Song}}, \bibinfo {author} {\bibfnamefont {S.}~\bibnamefont {Qin}}, \bibinfo {author} {\bibfnamefont {Q.}~\bibnamefont {Wen}},\ and\ \bibinfo {author} {\bibfnamefont {F.}~\bibnamefont {Gao}},\ }\bibfield  {title} {\bibinfo {title} {Quantum-assisted hierarchical fuzzy neural network for image classification},\ }\href {https://doi.org/10.1109/TFUZZ.2024.3435792} {\bibfield  {journal} {\bibinfo  {journal} {IEEE Trans. Fuzzy Syst.}\ }\textbf {\bibinfo {volume} {33}},\ \bibinfo {pages} {491} (\bibinfo {year} {2025}{\natexlab{a}})}\BibitemShut {NoStop}%
\bibitem [{\citenamefont {Bondarenko}\ and\ \citenamefont {Feldmann}(2020)}]{bondarenko2020quantum}%
  \BibitemOpen
  \bibfield  {author} {\bibinfo {author} {\bibfnamefont {D.}~\bibnamefont {Bondarenko}}\ and\ \bibinfo {author} {\bibfnamefont {P.}~\bibnamefont {Feldmann}},\ }\bibfield  {title} {\bibinfo {title} {Quantum autoencoders to denoise quantum data},\ }\href {https://doi.org/10.1103/PhysRevLett.124.130502} {\bibfield  {journal} {\bibinfo  {journal} {Phys. Rev. Lett.}\ }\textbf {\bibinfo {volume} {124}},\ \bibinfo {pages} {130502} (\bibinfo {year} {2020})}\BibitemShut {NoStop}%
\bibitem [{\citenamefont {Kerenidis}\ and\ \citenamefont {Landman}(2021)}]{kerenidis2021quantum}%
  \BibitemOpen
  \bibfield  {author} {\bibinfo {author} {\bibfnamefont {I.}~\bibnamefont {Kerenidis}}\ and\ \bibinfo {author} {\bibfnamefont {J.}~\bibnamefont {Landman}},\ }\bibfield  {title} {\bibinfo {title} {Quantum spectral clustering},\ }\href@noop {} {\bibfield  {journal} {\bibinfo  {journal} {Phys. Rev. A}\ }\textbf {\bibinfo {volume} {103}},\ \bibinfo {pages} {042415} (\bibinfo {year} {2021})}\BibitemShut {NoStop}%
\bibitem [{\citenamefont {Li}\ \emph {et~al.}(2025)\citenamefont {Li}, \citenamefont {Li}, \citenamefont {Song}, \citenamefont {Qin}, \citenamefont {Wen},\ and\ \citenamefont {Gao}}]{li2025efficient}%
  \BibitemOpen
  \bibfield  {author} {\bibinfo {author} {\bibfnamefont {L.}~\bibnamefont {Li}}, \bibinfo {author} {\bibfnamefont {J.}~\bibnamefont {Li}}, \bibinfo {author} {\bibfnamefont {Y.}~\bibnamefont {Song}}, \bibinfo {author} {\bibfnamefont {S.}~\bibnamefont {Qin}}, \bibinfo {author} {\bibfnamefont {Q.}~\bibnamefont {Wen}},\ and\ \bibinfo {author} {\bibfnamefont {F.}~\bibnamefont {Gao}},\ }\bibfield  {title} {\bibinfo {title} {An efficient quantum proactive incremental learning algorithm},\ }\href {https://doi.org/10.1007/s11433-024-2501-4} {\bibfield  {journal} {\bibinfo  {journal} {Sci. China-Phys. Mech. Astron.}\ }\textbf {\bibinfo {volume} {68}},\ \bibinfo {pages} {1} (\bibinfo {year} {2025})}\BibitemShut {NoStop}%
\bibitem [{\citenamefont {Wu}\ \emph {et~al.}(2025{\natexlab{b}})\citenamefont {Wu}, \citenamefont {Song}, \citenamefont {Li}, \citenamefont {Qin}, \citenamefont {Wen},\ and\ \citenamefont {Gao}}]{wu2025resource}%
  \BibitemOpen
  \bibfield  {author} {\bibinfo {author} {\bibfnamefont {S.-Y.}\ \bibnamefont {Wu}}, \bibinfo {author} {\bibfnamefont {Y.-Q.}\ \bibnamefont {Song}}, \bibinfo {author} {\bibfnamefont {R.-Z.}\ \bibnamefont {Li}}, \bibinfo {author} {\bibfnamefont {S.-J.}\ \bibnamefont {Qin}}, \bibinfo {author} {\bibfnamefont {Q.-Y.}\ \bibnamefont {Wen}},\ and\ \bibinfo {author} {\bibfnamefont {F.}~\bibnamefont {Gao}},\ }\bibfield  {title} {\bibinfo {title} {Resource-efficient adaptive variational quantum algorithm for combinatorial optimization problems},\ }\href {https://doi.org/10.1002/qute.202400484} {\bibfield  {journal} {\bibinfo  {journal} {Adv. Quantum Technol.}\ ,\ \bibinfo {pages} {2400484}} (\bibinfo {year} {2025}{\natexlab{b}})}\BibitemShut {NoStop}%
\bibitem [{\citenamefont {Havl{\'\i}{\v{c}}ek}\ \emph {et~al.}(2019)\citenamefont {Havl{\'\i}{\v{c}}ek}, \citenamefont {C{\'o}rcoles}, \citenamefont {Temme}, \citenamefont {Harrow}, \citenamefont {Kandala}, \citenamefont {Chow},\ and\ \citenamefont {Gambetta}}]{havlivcek2019supervised}%
  \BibitemOpen
  \bibfield  {author} {\bibinfo {author} {\bibfnamefont {V.}~\bibnamefont {Havl{\'\i}{\v{c}}ek}}, \bibinfo {author} {\bibfnamefont {A.~D.}\ \bibnamefont {C{\'o}rcoles}}, \bibinfo {author} {\bibfnamefont {K.}~\bibnamefont {Temme}}, \bibinfo {author} {\bibfnamefont {A.~W.}\ \bibnamefont {Harrow}}, \bibinfo {author} {\bibfnamefont {A.}~\bibnamefont {Kandala}}, \bibinfo {author} {\bibfnamefont {J.~M.}\ \bibnamefont {Chow}},\ and\ \bibinfo {author} {\bibfnamefont {J.~M.}\ \bibnamefont {Gambetta}},\ }\bibfield  {title} {\bibinfo {title} {Supervised learning with quantum-enhanced feature spaces},\ }\href {https://doi.org/10.1038/s41586-019-0980-2} {\bibfield  {journal} {\bibinfo  {journal} {Nature}\ }\textbf {\bibinfo {volume} {567}},\ \bibinfo {pages} {209} (\bibinfo {year} {2019})}\BibitemShut {NoStop}%
\bibitem [{\citenamefont {Schuld}\ and\ \citenamefont {Killoran}(2019)}]{schuld2019quantum}%
  \BibitemOpen
  \bibfield  {author} {\bibinfo {author} {\bibfnamefont {M.}~\bibnamefont {Schuld}}\ and\ \bibinfo {author} {\bibfnamefont {N.}~\bibnamefont {Killoran}},\ }\bibfield  {title} {\bibinfo {title} {Quantum machine learning in feature hilbert spaces},\ }\href {https://doi.org/10.1103/PhysRevLett.122.040504} {\bibfield  {journal} {\bibinfo  {journal} {Phys. Rev. Lett.}\ }\textbf {\bibinfo {volume} {122}},\ \bibinfo {pages} {040504} (\bibinfo {year} {2019})}\BibitemShut {NoStop}%
\bibitem [{\citenamefont {Schuld}(2021)}]{schuld2021supervised}%
  \BibitemOpen
  \bibfield  {author} {\bibinfo {author} {\bibfnamefont {M.}~\bibnamefont {Schuld}},\ }\bibfield  {title} {\bibinfo {title} {Supervised quantum machine learning models are kernel methods},\ }\bibfield  {journal} {\bibinfo  {journal} {arXiv preprint arXiv:2101.11020}\ }\href {https://doi.org/10.48550/arXiv.2101.11020} {10.48550/arXiv.2101.11020} (\bibinfo {year} {2021})\BibitemShut {NoStop}%
\bibitem [{\citenamefont {Henry}\ \emph {et~al.}(2021)\citenamefont {Henry}, \citenamefont {Thabet}, \citenamefont {Dalyac},\ and\ \citenamefont {Henriet}}]{henry2021quantum}%
  \BibitemOpen
  \bibfield  {author} {\bibinfo {author} {\bibfnamefont {L.-P.}\ \bibnamefont {Henry}}, \bibinfo {author} {\bibfnamefont {S.}~\bibnamefont {Thabet}}, \bibinfo {author} {\bibfnamefont {C.}~\bibnamefont {Dalyac}},\ and\ \bibinfo {author} {\bibfnamefont {L.}~\bibnamefont {Henriet}},\ }\bibfield  {title} {\bibinfo {title} {Quantum evolution kernel: Machine learning on graphs with programmable arrays of qubits},\ }\href {https://doi.org/10.1103/PhysRevA.104.032416} {\bibfield  {journal} {\bibinfo  {journal} {Phys. Rev. A}\ }\textbf {\bibinfo {volume} {104}},\ \bibinfo {pages} {032416} (\bibinfo {year} {2021})}\BibitemShut {NoStop}%
\bibitem [{\citenamefont {Liu}\ \emph {et~al.}(2022)\citenamefont {Liu}, \citenamefont {Tacchino}, \citenamefont {Glick}, \citenamefont {Jiang},\ and\ \citenamefont {Mezzacapo}}]{liu2022representation}%
  \BibitemOpen
  \bibfield  {author} {\bibinfo {author} {\bibfnamefont {J.}~\bibnamefont {Liu}}, \bibinfo {author} {\bibfnamefont {F.}~\bibnamefont {Tacchino}}, \bibinfo {author} {\bibfnamefont {J.~R.}\ \bibnamefont {Glick}}, \bibinfo {author} {\bibfnamefont {L.}~\bibnamefont {Jiang}},\ and\ \bibinfo {author} {\bibfnamefont {A.}~\bibnamefont {Mezzacapo}},\ }\bibfield  {title} {\bibinfo {title} {Representation learning via quantum neural tangent kernels},\ }\href {https://doi.org/10.1103/PRXQuantum.3.030323} {\bibfield  {journal} {\bibinfo  {journal} {PRX Quantum}\ }\textbf {\bibinfo {volume} {3}},\ \bibinfo {pages} {030323} (\bibinfo {year} {2022})}\BibitemShut {NoStop}%
\bibitem [{\citenamefont {Incudini}\ \emph {et~al.}(2023)\citenamefont {Incudini}, \citenamefont {Grossi}, \citenamefont {Mandarino}, \citenamefont {Vallecorsa}, \citenamefont {Pierro},\ and\ \citenamefont {Windridge}}]{incudini2023quantum}%
  \BibitemOpen
  \bibfield  {author} {\bibinfo {author} {\bibfnamefont {M.}~\bibnamefont {Incudini}}, \bibinfo {author} {\bibfnamefont {M.}~\bibnamefont {Grossi}}, \bibinfo {author} {\bibfnamefont {A.}~\bibnamefont {Mandarino}}, \bibinfo {author} {\bibfnamefont {S.}~\bibnamefont {Vallecorsa}}, \bibinfo {author} {\bibfnamefont {A.~D.}\ \bibnamefont {Pierro}},\ and\ \bibinfo {author} {\bibfnamefont {D.}~\bibnamefont {Windridge}},\ }\bibfield  {title} {\bibinfo {title} {The quantum path kernel: A generalized neural tangent kernel for deep quantum machine learning},\ }\href {https://doi.org/10.1109/TQE.2023.3287736} {\bibfield  {journal} {\bibinfo  {journal} {IEEE Trans. Quantum Eng.}\ }\textbf {\bibinfo {volume} {4}},\ \bibinfo {pages} {1} (\bibinfo {year} {2023})}\BibitemShut {NoStop}%
\bibitem [{\citenamefont {Shirai}\ \emph {et~al.}(2024)\citenamefont {Shirai}, \citenamefont {Kubo}, \citenamefont {Mitarai},\ and\ \citenamefont {Fujii}}]{shirai2024quantum}%
  \BibitemOpen
  \bibfield  {author} {\bibinfo {author} {\bibfnamefont {N.}~\bibnamefont {Shirai}}, \bibinfo {author} {\bibfnamefont {K.}~\bibnamefont {Kubo}}, \bibinfo {author} {\bibfnamefont {K.}~\bibnamefont {Mitarai}},\ and\ \bibinfo {author} {\bibfnamefont {K.}~\bibnamefont {Fujii}},\ }\bibfield  {title} {\bibinfo {title} {Quantum tangent kernel},\ }\href {https://doi.org/10.1103/PhysRevResearch.6.033179} {\bibfield  {journal} {\bibinfo  {journal} {Physical Review Research}\ }\textbf {\bibinfo {volume} {6}},\ \bibinfo {pages} {033179} (\bibinfo {year} {2024})}\BibitemShut {NoStop}%
\bibitem [{\citenamefont {Liu}\ \emph {et~al.}(2021)\citenamefont {Liu}, \citenamefont {Arunachalam},\ and\ \citenamefont {Temme}}]{liu2021rigorous}%
  \BibitemOpen
  \bibfield  {author} {\bibinfo {author} {\bibfnamefont {Y.}~\bibnamefont {Liu}}, \bibinfo {author} {\bibfnamefont {S.}~\bibnamefont {Arunachalam}},\ and\ \bibinfo {author} {\bibfnamefont {K.}~\bibnamefont {Temme}},\ }\bibfield  {title} {\bibinfo {title} {A rigorous and robust quantum speed-up in supervised machine learning},\ }\href {https://doi.org/10.1038/s41567-021-01287-z} {\bibfield  {journal} {\bibinfo  {journal} {Nat. Phys.}\ }\textbf {\bibinfo {volume} {17}},\ \bibinfo {pages} {1013} (\bibinfo {year} {2021})}\BibitemShut {NoStop}%
\bibitem [{\citenamefont {Wang}\ \emph {et~al.}(2021)\citenamefont {Wang}, \citenamefont {Du}, \citenamefont {Luo},\ and\ \citenamefont {Tao}}]{wang2021towards}%
  \BibitemOpen
  \bibfield  {author} {\bibinfo {author} {\bibfnamefont {X.}~\bibnamefont {Wang}}, \bibinfo {author} {\bibfnamefont {Y.}~\bibnamefont {Du}}, \bibinfo {author} {\bibfnamefont {Y.}~\bibnamefont {Luo}},\ and\ \bibinfo {author} {\bibfnamefont {D.}~\bibnamefont {Tao}},\ }\bibfield  {title} {\bibinfo {title} {Towards understanding the power of quantum kernels in the nisq era},\ }\href {https://doi.org/10.22331/q-2021-08-30-531} {\bibfield  {journal} {\bibinfo  {journal} {Quantum}\ }\textbf {\bibinfo {volume} {5}},\ \bibinfo {pages} {531} (\bibinfo {year} {2021})}\BibitemShut {NoStop}%
\bibitem [{\citenamefont {Hubregtsen}\ \emph {et~al.}(2022)\citenamefont {Hubregtsen}, \citenamefont {Wierichs}, \citenamefont {Gil-Fuster}, \citenamefont {Derks}, \citenamefont {Faehrmann},\ and\ \citenamefont {Meyer}}]{hubregtsen2022training}%
  \BibitemOpen
  \bibfield  {author} {\bibinfo {author} {\bibfnamefont {T.}~\bibnamefont {Hubregtsen}}, \bibinfo {author} {\bibfnamefont {D.}~\bibnamefont {Wierichs}}, \bibinfo {author} {\bibfnamefont {E.}~\bibnamefont {Gil-Fuster}}, \bibinfo {author} {\bibfnamefont {P.-J.~H.}\ \bibnamefont {Derks}}, \bibinfo {author} {\bibfnamefont {P.~K.}\ \bibnamefont {Faehrmann}},\ and\ \bibinfo {author} {\bibfnamefont {J.~J.}\ \bibnamefont {Meyer}},\ }\bibfield  {title} {\bibinfo {title} {Training quantum embedding kernels on near-term quantum computers},\ }\href {https://doi.org/10.1103/PhysRevA.106.042431} {\bibfield  {journal} {\bibinfo  {journal} {Phys. Rev. A}\ }\textbf {\bibinfo {volume} {106}},\ \bibinfo {pages} {042431} (\bibinfo {year} {2022})}\BibitemShut {NoStop}%
\bibitem [{\citenamefont {Thanasilp}\ \emph {et~al.}(2024)\citenamefont {Thanasilp}, \citenamefont {Wang}, \citenamefont {Cerezo},\ and\ \citenamefont {Holmes}}]{thanasilp2024exponential}%
  \BibitemOpen
  \bibfield  {author} {\bibinfo {author} {\bibfnamefont {S.}~\bibnamefont {Thanasilp}}, \bibinfo {author} {\bibfnamefont {S.}~\bibnamefont {Wang}}, \bibinfo {author} {\bibfnamefont {M.}~\bibnamefont {Cerezo}},\ and\ \bibinfo {author} {\bibfnamefont {Z.}~\bibnamefont {Holmes}},\ }\bibfield  {title} {\bibinfo {title} {Exponential concentration in quantum kernel methods},\ }\href {https://doi.org/10.1038/s41467-024-49287-w} {\bibfield  {journal} {\bibinfo  {journal} {Nat. Commun.}\ }\textbf {\bibinfo {volume} {15}},\ \bibinfo {pages} {5200} (\bibinfo {year} {2024})}\BibitemShut {NoStop}%
\bibitem [{\citenamefont {Miyabe}\ \emph {et~al.}(2023)\citenamefont {Miyabe}, \citenamefont {Quanz}, \citenamefont {Shimada}, \citenamefont {Mitra}, \citenamefont {Yamamoto}, \citenamefont {Rastunkov}, \citenamefont {Alevras}, \citenamefont {Metcalf}, \citenamefont {King}, \citenamefont {Mamouei}, \citenamefont {Jackson}, \citenamefont {Brown}, \citenamefont {Intallura},\ and\ \citenamefont {Park}}]{miyabe2023quantum}%
  \BibitemOpen
  \bibfield  {author} {\bibinfo {author} {\bibfnamefont {S.}~\bibnamefont {Miyabe}}, \bibinfo {author} {\bibfnamefont {B.}~\bibnamefont {Quanz}}, \bibinfo {author} {\bibfnamefont {N.}~\bibnamefont {Shimada}}, \bibinfo {author} {\bibfnamefont {A.}~\bibnamefont {Mitra}}, \bibinfo {author} {\bibfnamefont {T.}~\bibnamefont {Yamamoto}}, \bibinfo {author} {\bibfnamefont {V.}~\bibnamefont {Rastunkov}}, \bibinfo {author} {\bibfnamefont {D.}~\bibnamefont {Alevras}}, \bibinfo {author} {\bibfnamefont {M.}~\bibnamefont {Metcalf}}, \bibinfo {author} {\bibfnamefont {D.~J.~M.}\ \bibnamefont {King}}, \bibinfo {author} {\bibfnamefont {M.}~\bibnamefont {Mamouei}}, \bibinfo {author} {\bibfnamefont {M.~D.}\ \bibnamefont {Jackson}}, \bibinfo {author} {\bibfnamefont {M.}~\bibnamefont {Brown}}, \bibinfo {author} {\bibfnamefont {P.}~\bibnamefont {Intallura}},\ and\ \bibinfo {author} {\bibfnamefont {J.-E.}\ \bibnamefont {Park}},\ }\href {https://arxiv.org/abs/2312.00260} {\bibinfo {title} {Quantum multiple kernel learning in
  financial classification tasks}} (\bibinfo {year} {2023}),\ \Eprint {https://arxiv.org/abs/2312.00260} {arXiv:2312.00260 [quant-ph]} \BibitemShut {NoStop}%
\bibitem [{\citenamefont {Fu}\ \emph {et~al.}(2024)\citenamefont {Fu}, \citenamefont {Zhang}, \citenamefont {Yang},\ and\ \citenamefont {Qi}}]{fu2024exploiting}%
  \BibitemOpen
  \bibfield  {author} {\bibinfo {author} {\bibfnamefont {X.}~\bibnamefont {Fu}}, \bibinfo {author} {\bibfnamefont {X.-L.}\ \bibnamefont {Zhang}}, \bibinfo {author} {\bibfnamefont {C.-H.~H.}\ \bibnamefont {Yang}},\ and\ \bibinfo {author} {\bibfnamefont {J.}~\bibnamefont {Qi}},\ }\bibfield  {title} {\bibinfo {title} {Exploiting a quantum multiple kernel learning approach for low-resource spoken command recognition},\ }in\ \href {https://doi.org/10.1109/ICASSP48485.2024.10448120} {\emph {\bibinfo {booktitle} {ICASSP 2024 - 2024 IEEE International Conference on Acoustics, Speech and Signal Processing (ICASSP)}}}\ (\bibinfo {year} {2024})\ pp.\ \bibinfo {pages} {12931--12935}\BibitemShut {NoStop}%
\bibitem [{\citenamefont {Vedaie}\ \emph {et~al.}(2020)\citenamefont {Vedaie}, \citenamefont {Noori}, \citenamefont {Oberoi}, \citenamefont {Sanders},\ and\ \citenamefont {Zahedinejad}}]{vedaie2020quantum}%
  \BibitemOpen
  \bibfield  {author} {\bibinfo {author} {\bibfnamefont {S.~S.}\ \bibnamefont {Vedaie}}, \bibinfo {author} {\bibfnamefont {M.}~\bibnamefont {Noori}}, \bibinfo {author} {\bibfnamefont {J.~S.}\ \bibnamefont {Oberoi}}, \bibinfo {author} {\bibfnamefont {B.~C.}\ \bibnamefont {Sanders}},\ and\ \bibinfo {author} {\bibfnamefont {E.}~\bibnamefont {Zahedinejad}},\ }\href {https://arxiv.org/abs/2011.09694} {\bibinfo {title} {Quantum multiple kernel learning}} (\bibinfo {year} {2020}),\ \Eprint {https://arxiv.org/abs/2011.09694} {arXiv:2011.09694 [quant-ph]} \BibitemShut {NoStop}%
\bibitem [{\citenamefont {Ghukasyan}\ \emph {et~al.}(2023)\citenamefont {Ghukasyan}, \citenamefont {Baker}, \citenamefont {Goktas}, \citenamefont {Carrasquilla},\ and\ \citenamefont {Radha}}]{ghukasyan2023quantum}%
  \BibitemOpen
  \bibfield  {author} {\bibinfo {author} {\bibfnamefont {A.}~\bibnamefont {Ghukasyan}}, \bibinfo {author} {\bibfnamefont {J.~S.}\ \bibnamefont {Baker}}, \bibinfo {author} {\bibfnamefont {O.}~\bibnamefont {Goktas}}, \bibinfo {author} {\bibfnamefont {J.}~\bibnamefont {Carrasquilla}},\ and\ \bibinfo {author} {\bibfnamefont {S.~K.}\ \bibnamefont {Radha}},\ }\href {https://arxiv.org/abs/2305.17707} {\bibinfo {title} {Quantum-classical multiple kernel learning}} (\bibinfo {year} {2023}),\ \Eprint {https://arxiv.org/abs/2305.17707} {arXiv:2305.17707 [quant-ph]} \BibitemShut {NoStop}%
\bibitem [{\citenamefont {Zhao}\ \emph {et~al.}(2017)\citenamefont {Zhao}, \citenamefont {Xie}, \citenamefont {Xu},\ and\ \citenamefont {Sun}}]{zhao2017multi}%
  \BibitemOpen
  \bibfield  {author} {\bibinfo {author} {\bibfnamefont {J.}~\bibnamefont {Zhao}}, \bibinfo {author} {\bibfnamefont {X.}~\bibnamefont {Xie}}, \bibinfo {author} {\bibfnamefont {X.}~\bibnamefont {Xu}},\ and\ \bibinfo {author} {\bibfnamefont {S.}~\bibnamefont {Sun}},\ }\bibfield  {title} {\bibinfo {title} {Multi-view learning overview: Recent progress and new challenges},\ }\href {https://doi.org/10.1016/j.inffus.2017.02.007} {\bibfield  {journal} {\bibinfo  {journal} {Inf. Fusion}\ }\textbf {\bibinfo {volume} {38}},\ \bibinfo {pages} {43} (\bibinfo {year} {2017})}\BibitemShut {NoStop}%
\bibitem [{\citenamefont {Yu}\ \emph {et~al.}(2025)\citenamefont {Yu}, \citenamefont {Dong}, \citenamefont {Yu}, \citenamefont {Yang}, \citenamefont {Fan},\ and\ \citenamefont {Chen}}]{yu2025review}%
  \BibitemOpen
  \bibfield  {author} {\bibinfo {author} {\bibfnamefont {Z.}~\bibnamefont {Yu}}, \bibinfo {author} {\bibfnamefont {Z.}~\bibnamefont {Dong}}, \bibinfo {author} {\bibfnamefont {C.}~\bibnamefont {Yu}}, \bibinfo {author} {\bibfnamefont {K.}~\bibnamefont {Yang}}, \bibinfo {author} {\bibfnamefont {Z.}~\bibnamefont {Fan}},\ and\ \bibinfo {author} {\bibfnamefont {C.~P.}\ \bibnamefont {Chen}},\ }\bibfield  {title} {\bibinfo {title} {A review on multi-view learning},\ }\href {https://doi.org/10.1007/s11704-024-40004-w} {\bibfield  {journal} {\bibinfo  {journal} {Front.. Comput. Sci.}\ }\textbf {\bibinfo {volume} {19}},\ \bibinfo {pages} {197334} (\bibinfo {year} {2025})}\BibitemShut {NoStop}%
\bibitem [{\citenamefont {Yeh}\ \emph {et~al.}(2012)\citenamefont {Yeh}, \citenamefont {Lin}, \citenamefont {Chung},\ and\ \citenamefont {Wang}}]{yeh2012novel}%
  \BibitemOpen
  \bibfield  {author} {\bibinfo {author} {\bibfnamefont {Y.-R.}\ \bibnamefont {Yeh}}, \bibinfo {author} {\bibfnamefont {T.-C.}\ \bibnamefont {Lin}}, \bibinfo {author} {\bibfnamefont {Y.-Y.}\ \bibnamefont {Chung}},\ and\ \bibinfo {author} {\bibfnamefont {Y.-C.~F.}\ \bibnamefont {Wang}},\ }\bibfield  {title} {\bibinfo {title} {A novel multiple kernel learning framework for heterogeneous feature fusion and variable selection},\ }\href {https://doi.org/10.1109/TMM.2012.2188783} {\bibfield  {journal} {\bibinfo  {journal} {IEEE Trans. Multimedia}\ }\textbf {\bibinfo {volume} {14}},\ \bibinfo {pages} {563} (\bibinfo {year} {2012})}\BibitemShut {NoStop}%
\bibitem [{\citenamefont {Chen}\ \emph {et~al.}(2014)\citenamefont {Chen}, \citenamefont {Chen}, \citenamefont {Chi},\ and\ \citenamefont {Fu}}]{chen2014emotion}%
  \BibitemOpen
  \bibfield  {author} {\bibinfo {author} {\bibfnamefont {J.}~\bibnamefont {Chen}}, \bibinfo {author} {\bibfnamefont {Z.}~\bibnamefont {Chen}}, \bibinfo {author} {\bibfnamefont {Z.}~\bibnamefont {Chi}},\ and\ \bibinfo {author} {\bibfnamefont {H.}~\bibnamefont {Fu}},\ }\bibfield  {title} {\bibinfo {title} {Emotion recognition in the wild with feature fusion and multiple kernel learning},\ }in\ \href {https://doi.org/10.1145/2663204.266627} {\emph {\bibinfo {booktitle} {Proceedings of the 16th International Conference on Multimodal Interaction}}}\ (\bibinfo {year} {2014})\ pp.\ \bibinfo {pages} {508--513}\BibitemShut {NoStop}%
\bibitem [{\citenamefont {Yan}\ \emph {et~al.}(2023)\citenamefont {Yan}, \citenamefont {Li},\ and\ \citenamefont {Yang}}]{yan2023towards}%
  \BibitemOpen
  \bibfield  {author} {\bibinfo {author} {\bibfnamefont {W.}~\bibnamefont {Yan}}, \bibinfo {author} {\bibfnamefont {Y.}~\bibnamefont {Li}},\ and\ \bibinfo {author} {\bibfnamefont {M.}~\bibnamefont {Yang}},\ }\bibfield  {title} {\bibinfo {title} {Towards deeper match for multi-view oriented multiple kernel learning},\ }\href {https://doi.org/10.1016/j.patcog.2022.109119} {\bibfield  {journal} {\bibinfo  {journal} {Pattern Recognit.}\ }\textbf {\bibinfo {volume} {134}},\ \bibinfo {pages} {109119} (\bibinfo {year} {2023})}\BibitemShut {NoStop}%
\bibitem [{\citenamefont {Wang}\ \emph {et~al.}(2017)\citenamefont {Wang}, \citenamefont {Liu}, \citenamefont {Dou}, \citenamefont {Lv},\ and\ \citenamefont {Lu}}]{wang2017multiple}%
  \BibitemOpen
  \bibfield  {author} {\bibinfo {author} {\bibfnamefont {Y.}~\bibnamefont {Wang}}, \bibinfo {author} {\bibfnamefont {X.}~\bibnamefont {Liu}}, \bibinfo {author} {\bibfnamefont {Y.}~\bibnamefont {Dou}}, \bibinfo {author} {\bibfnamefont {Q.}~\bibnamefont {Lv}},\ and\ \bibinfo {author} {\bibfnamefont {Y.}~\bibnamefont {Lu}},\ }\bibfield  {title} {\bibinfo {title} {Multiple kernel learning with hybrid kernel alignment maximization},\ }\href {https://doi.org/10.1016/j.patcog.2017.05.005} {\bibfield  {journal} {\bibinfo  {journal} {Pattern Recognit.}\ }\textbf {\bibinfo {volume} {70}},\ \bibinfo {pages} {104} (\bibinfo {year} {2017})}\BibitemShut {NoStop}%
\bibitem [{\citenamefont {Wang}\ \emph {et~al.}(2018)\citenamefont {Wang}, \citenamefont {Dou}, \citenamefont {Liu}, \citenamefont {Xia}, \citenamefont {Lv},\ and\ \citenamefont {Yang}}]{wang2018local}%
  \BibitemOpen
  \bibfield  {author} {\bibinfo {author} {\bibfnamefont {Q.}~\bibnamefont {Wang}}, \bibinfo {author} {\bibfnamefont {Y.}~\bibnamefont {Dou}}, \bibinfo {author} {\bibfnamefont {X.}~\bibnamefont {Liu}}, \bibinfo {author} {\bibfnamefont {F.}~\bibnamefont {Xia}}, \bibinfo {author} {\bibfnamefont {Q.}~\bibnamefont {Lv}},\ and\ \bibinfo {author} {\bibfnamefont {K.}~\bibnamefont {Yang}},\ }\bibfield  {title} {\bibinfo {title} {Local kernel alignment based multi-view clustering using extreme learning machine},\ }\href {https://doi.org/10.1016/j.neucom.2017.09.060} {\bibfield  {journal} {\bibinfo  {journal} {Neurocomputing}\ }\textbf {\bibinfo {volume} {275}},\ \bibinfo {pages} {1099} (\bibinfo {year} {2018})}\BibitemShut {NoStop}%
\bibitem [{\citenamefont {Zhang}\ \emph {et~al.}(2021)\citenamefont {Zhang}, \citenamefont {Liu}, \citenamefont {Gong}, \citenamefont {Wang}, \citenamefont {Niu},\ and\ \citenamefont {Shen}}]{zhang2021late}%
  \BibitemOpen
  \bibfield  {author} {\bibinfo {author} {\bibfnamefont {T.}~\bibnamefont {Zhang}}, \bibinfo {author} {\bibfnamefont {X.}~\bibnamefont {Liu}}, \bibinfo {author} {\bibfnamefont {L.}~\bibnamefont {Gong}}, \bibinfo {author} {\bibfnamefont {S.}~\bibnamefont {Wang}}, \bibinfo {author} {\bibfnamefont {X.}~\bibnamefont {Niu}},\ and\ \bibinfo {author} {\bibfnamefont {L.}~\bibnamefont {Shen}},\ }\bibfield  {title} {\bibinfo {title} {Late fusion multiple kernel clustering with local kernel alignment maximization},\ }\href {https://doi.org/10.1109/TMM.2021.3136094} {\bibfield  {journal} {\bibinfo  {journal} {IEEE Trans. Multimedia}\ }\textbf {\bibinfo {volume} {25}},\ \bibinfo {pages} {993} (\bibinfo {year} {2021})}\BibitemShut {NoStop}%
\bibitem [{\citenamefont {Kandala}\ \emph {et~al.}(2017)\citenamefont {Kandala}, \citenamefont {Mezzacapo}, \citenamefont {Temme}, \citenamefont {Takita}, \citenamefont {Brink}, \citenamefont {Chow},\ and\ \citenamefont {Gambetta}}]{kandala2017hardware}%
  \BibitemOpen
  \bibfield  {author} {\bibinfo {author} {\bibfnamefont {A.}~\bibnamefont {Kandala}}, \bibinfo {author} {\bibfnamefont {A.}~\bibnamefont {Mezzacapo}}, \bibinfo {author} {\bibfnamefont {K.}~\bibnamefont {Temme}}, \bibinfo {author} {\bibfnamefont {M.}~\bibnamefont {Takita}}, \bibinfo {author} {\bibfnamefont {M.}~\bibnamefont {Brink}}, \bibinfo {author} {\bibfnamefont {J.~M.}\ \bibnamefont {Chow}},\ and\ \bibinfo {author} {\bibfnamefont {J.~M.}\ \bibnamefont {Gambetta}},\ }\bibfield  {title} {\bibinfo {title} {Hardware-efficient variational quantum eigensolver for small molecules and quantum magnets},\ }\href {https://doi.org/10.1038/nature23879} {\bibfield  {journal} {\bibinfo  {journal} {Nature}\ }\textbf {\bibinfo {volume} {549}},\ \bibinfo {pages} {242} (\bibinfo {year} {2017})}\BibitemShut {NoStop}%
\bibitem [{\citenamefont {Farhi}\ \emph {et~al.}(2014)\citenamefont {Farhi}, \citenamefont {Goldstone},\ and\ \citenamefont {Gutmann}}]{farhi2014quantum}%
  \BibitemOpen
  \bibfield  {author} {\bibinfo {author} {\bibfnamefont {E.}~\bibnamefont {Farhi}}, \bibinfo {author} {\bibfnamefont {J.}~\bibnamefont {Goldstone}},\ and\ \bibinfo {author} {\bibfnamefont {S.}~\bibnamefont {Gutmann}},\ }\href {https://arxiv.org/abs/1411.4028} {\bibinfo {title} {A quantum approximate optimization algorithm}} (\bibinfo {year} {2014}),\ \Eprint {https://arxiv.org/abs/1411.4028} {arXiv:1411.4028 [quant-ph]} \BibitemShut {NoStop}%
\bibitem [{\citenamefont {Hadfield}\ \emph {et~al.}(2019)\citenamefont {Hadfield}, \citenamefont {Wang}, \citenamefont {O’gorman}, \citenamefont {Rieffel}, \citenamefont {Venturelli},\ and\ \citenamefont {Biswas}}]{hadfield2019quantum}%
  \BibitemOpen
  \bibfield  {author} {\bibinfo {author} {\bibfnamefont {S.}~\bibnamefont {Hadfield}}, \bibinfo {author} {\bibfnamefont {Z.}~\bibnamefont {Wang}}, \bibinfo {author} {\bibfnamefont {B.}~\bibnamefont {O’gorman}}, \bibinfo {author} {\bibfnamefont {E.~G.}\ \bibnamefont {Rieffel}}, \bibinfo {author} {\bibfnamefont {D.}~\bibnamefont {Venturelli}},\ and\ \bibinfo {author} {\bibfnamefont {R.}~\bibnamefont {Biswas}},\ }\bibfield  {title} {\bibinfo {title} {From the quantum approximate optimization algorithm to a quantum alternating operator ansatz},\ }\href {https://doi.org/10.3390/a12020034} {\bibfield  {journal} {\bibinfo  {journal} {Algorithms}\ }\textbf {\bibinfo {volume} {12}},\ \bibinfo {pages} {34} (\bibinfo {year} {2019})}\BibitemShut {NoStop}%
\bibitem [{\citenamefont {Suykens}\ and\ \citenamefont {Vandewalle}(1999)}]{suykens1999least}%
  \BibitemOpen
  \bibfield  {author} {\bibinfo {author} {\bibfnamefont {J.~A.}\ \bibnamefont {Suykens}}\ and\ \bibinfo {author} {\bibfnamefont {J.}~\bibnamefont {Vandewalle}},\ }\bibfield  {title} {\bibinfo {title} {Least squares support vector machine classifiers},\ }\href {https://doi.org/10.1023/A:1018628609742} {\bibfield  {journal} {\bibinfo  {journal} {Neural Process. Lett.}\ }\textbf {\bibinfo {volume} {9}},\ \bibinfo {pages} {293} (\bibinfo {year} {1999})}\BibitemShut {NoStop}%
\bibitem [{\citenamefont {Wang}\ \emph {et~al.}(2015)\citenamefont {Wang}, \citenamefont {Zhao},\ and\ \citenamefont {Tian}}]{wang2015overview}%
  \BibitemOpen
  \bibfield  {author} {\bibinfo {author} {\bibfnamefont {T.}~\bibnamefont {Wang}}, \bibinfo {author} {\bibfnamefont {D.}~\bibnamefont {Zhao}},\ and\ \bibinfo {author} {\bibfnamefont {S.}~\bibnamefont {Tian}},\ }\bibfield  {title} {\bibinfo {title} {An overview of kernel alignment and its applications},\ }\href {https://doi.org/10.1007/s10462-012-9369-4} {\bibfield  {journal} {\bibinfo  {journal} {Artif. Intell. Rev.}\ }\textbf {\bibinfo {volume} {43}},\ \bibinfo {pages} {179} (\bibinfo {year} {2015})}\BibitemShut {NoStop}%
\bibitem [{\citenamefont {Ruder}(2017)}]{ruder2016overview}%
  \BibitemOpen
  \bibfield  {author} {\bibinfo {author} {\bibfnamefont {S.}~\bibnamefont {Ruder}},\ }\href {https://arxiv.org/abs/1609.04747} {\bibinfo {title} {An overview of gradient descent optimization algorithms}} (\bibinfo {year} {2017}),\ \Eprint {https://arxiv.org/abs/1609.04747} {arXiv:1609.04747 [cs.LG]} \BibitemShut {NoStop}%
\bibitem [{\citenamefont {Romero}\ \emph {et~al.}(2018)\citenamefont {Romero}, \citenamefont {Babbush}, \citenamefont {McClean}, \citenamefont {Hempel}, \citenamefont {Love},\ and\ \citenamefont {Aspuru-Guzik}}]{romero18strategies}%
  \BibitemOpen
  \bibfield  {author} {\bibinfo {author} {\bibfnamefont {J.}~\bibnamefont {Romero}}, \bibinfo {author} {\bibfnamefont {R.}~\bibnamefont {Babbush}}, \bibinfo {author} {\bibfnamefont {J.~R.}\ \bibnamefont {McClean}}, \bibinfo {author} {\bibfnamefont {C.}~\bibnamefont {Hempel}}, \bibinfo {author} {\bibfnamefont {P.~J.}\ \bibnamefont {Love}},\ and\ \bibinfo {author} {\bibfnamefont {A.}~\bibnamefont {Aspuru-Guzik}},\ }\bibfield  {title} {\bibinfo {title} {Strategies for quantum computing molecular energies using the unitary coupled cluster ansatz},\ }\href {https://doi.org/10.1088/2058-9565/aad3e4} {\bibfield  {journal} {\bibinfo  {journal} {Quantum Sci. Technol.}\ }\textbf {\bibinfo {volume} {4}},\ \bibinfo {pages} {014008} (\bibinfo {year} {2018})}\BibitemShut {NoStop}%
\bibitem [{\citenamefont {Schuld}\ \emph {et~al.}(2019)\citenamefont {Schuld}, \citenamefont {Bergholm}, \citenamefont {Gogolin}, \citenamefont {Izaac},\ and\ \citenamefont {Killoran}}]{schuld2019evaluating}%
  \BibitemOpen
  \bibfield  {author} {\bibinfo {author} {\bibfnamefont {M.}~\bibnamefont {Schuld}}, \bibinfo {author} {\bibfnamefont {V.}~\bibnamefont {Bergholm}}, \bibinfo {author} {\bibfnamefont {C.}~\bibnamefont {Gogolin}}, \bibinfo {author} {\bibfnamefont {J.}~\bibnamefont {Izaac}},\ and\ \bibinfo {author} {\bibfnamefont {N.}~\bibnamefont {Killoran}},\ }\bibfield  {title} {\bibinfo {title} {Evaluating analytic gradients on quantum hardware},\ }\href {https://doi.org/10.1103/PhysRevA.99.032331} {\bibfield  {journal} {\bibinfo  {journal} {Phys. Rev. A}\ }\textbf {\bibinfo {volume} {99}},\ \bibinfo {pages} {032331} (\bibinfo {year} {2019})}\BibitemShut {NoStop}%
\bibitem [{\citenamefont {Duin}(1998)}]{mfeat_dataset}%
  \BibitemOpen
  \bibfield  {author} {\bibinfo {author} {\bibfnamefont {R.}~\bibnamefont {Duin}},\ }\href {https://doi.org/10.24432/C5HC70} {\bibinfo {title} {{Multiple Features}}},\ \bibinfo {howpublished} {UCI Machine Learning Repository} (\bibinfo {year} {1998})\BibitemShut {NoStop}%
\bibitem [{\citenamefont {Bergholm}\ \emph {et~al.}(2022)\citenamefont {Bergholm}, \citenamefont {Izaac}, \citenamefont {Schuld}, \citenamefont {Gogolin}, \citenamefont {Ahmed}, \citenamefont {Ajith}, \citenamefont {Alam}, \citenamefont {Alonso-Linaje}, \citenamefont {AkashNarayanan}, \citenamefont {Asadi}, \citenamefont {Arrazola}, \citenamefont {Azad}, \citenamefont {Banning}, \citenamefont {Blank}, \citenamefont {Bromley}, \citenamefont {Cordier}, \citenamefont {Ceroni}, \citenamefont {Delgado}, \citenamefont {Matteo}, \citenamefont {Dusko}, \citenamefont {Garg}, \citenamefont {Guala}, \citenamefont {Hayes}, \citenamefont {Hill}, \citenamefont {Ijaz}, \citenamefont {Isacsson}, \citenamefont {Ittah}, \citenamefont {Jahangiri}, \citenamefont {Jain}, \citenamefont {Jiang}, \citenamefont {Khandelwal}, \citenamefont {Kottmann}, \citenamefont {Lang}, \citenamefont {Lee}, \citenamefont {Loke}, \citenamefont {Lowe}, \citenamefont {McKiernan}, \citenamefont {Meyer}, \citenamefont {Montañez-Barrera}, \citenamefont
  {Moyard}, \citenamefont {Niu}, \citenamefont {O'Riordan}, \citenamefont {Oud}, \citenamefont {Panigrahi}, \citenamefont {Park}, \citenamefont {Polatajko}, \citenamefont {Quesada}, \citenamefont {Roberts}, \citenamefont {Sá}, \citenamefont {Schoch}, \citenamefont {Shi}, \citenamefont {Shu}, \citenamefont {Sim}, \citenamefont {Singh}, \citenamefont {Strandberg}, \citenamefont {Soni}, \citenamefont {Száva}, \citenamefont {Thabet}, \citenamefont {Vargas-Hernández}, \citenamefont {Vincent}, \citenamefont {Vitucci}, \citenamefont {Weber}, \citenamefont {Wierichs}, \citenamefont {Wiersema}, \citenamefont {Willmann}, \citenamefont {Wong}, \citenamefont {Zhang},\ and\ \citenamefont {Killoran}}]{bergholm2018pennylane}%
  \BibitemOpen
  \bibfield  {author} {\bibinfo {author} {\bibfnamefont {V.}~\bibnamefont {Bergholm}}, \bibinfo {author} {\bibfnamefont {J.}~\bibnamefont {Izaac}}, \bibinfo {author} {\bibfnamefont {M.}~\bibnamefont {Schuld}}, \bibinfo {author} {\bibfnamefont {C.}~\bibnamefont {Gogolin}}, \bibinfo {author} {\bibfnamefont {S.}~\bibnamefont {Ahmed}}, \bibinfo {author} {\bibfnamefont {V.}~\bibnamefont {Ajith}}, \bibinfo {author} {\bibfnamefont {M.~S.}\ \bibnamefont {Alam}}, \bibinfo {author} {\bibfnamefont {G.}~\bibnamefont {Alonso-Linaje}}, \bibinfo {author} {\bibfnamefont {B.}~\bibnamefont {AkashNarayanan}}, \bibinfo {author} {\bibfnamefont {A.}~\bibnamefont {Asadi}}, \bibinfo {author} {\bibfnamefont {J.~M.}\ \bibnamefont {Arrazola}}, \bibinfo {author} {\bibfnamefont {U.}~\bibnamefont {Azad}}, \bibinfo {author} {\bibfnamefont {S.}~\bibnamefont {Banning}}, \bibinfo {author} {\bibfnamefont {C.}~\bibnamefont {Blank}}, \bibinfo {author} {\bibfnamefont {T.~R.}\ \bibnamefont {Bromley}}, \bibinfo {author} {\bibfnamefont {B.~A.}\
  \bibnamefont {Cordier}}, \bibinfo {author} {\bibfnamefont {J.}~\bibnamefont {Ceroni}}, \bibinfo {author} {\bibfnamefont {A.}~\bibnamefont {Delgado}}, \bibinfo {author} {\bibfnamefont {O.~D.}\ \bibnamefont {Matteo}}, \bibinfo {author} {\bibfnamefont {A.}~\bibnamefont {Dusko}}, \bibinfo {author} {\bibfnamefont {T.}~\bibnamefont {Garg}}, \bibinfo {author} {\bibfnamefont {D.}~\bibnamefont {Guala}}, \bibinfo {author} {\bibfnamefont {A.}~\bibnamefont {Hayes}}, \bibinfo {author} {\bibfnamefont {R.}~\bibnamefont {Hill}}, \bibinfo {author} {\bibfnamefont {A.}~\bibnamefont {Ijaz}}, \bibinfo {author} {\bibfnamefont {T.}~\bibnamefont {Isacsson}}, \bibinfo {author} {\bibfnamefont {D.}~\bibnamefont {Ittah}}, \bibinfo {author} {\bibfnamefont {S.}~\bibnamefont {Jahangiri}}, \bibinfo {author} {\bibfnamefont {P.}~\bibnamefont {Jain}}, \bibinfo {author} {\bibfnamefont {E.}~\bibnamefont {Jiang}}, \bibinfo {author} {\bibfnamefont {A.}~\bibnamefont {Khandelwal}}, \bibinfo {author} {\bibfnamefont {K.}~\bibnamefont {Kottmann}},
  \bibinfo {author} {\bibfnamefont {R.~A.}\ \bibnamefont {Lang}}, \bibinfo {author} {\bibfnamefont {C.}~\bibnamefont {Lee}}, \bibinfo {author} {\bibfnamefont {T.}~\bibnamefont {Loke}}, \bibinfo {author} {\bibfnamefont {A.}~\bibnamefont {Lowe}}, \bibinfo {author} {\bibfnamefont {K.}~\bibnamefont {McKiernan}}, \bibinfo {author} {\bibfnamefont {J.~J.}\ \bibnamefont {Meyer}}, \bibinfo {author} {\bibfnamefont {J.~A.}\ \bibnamefont {Montañez-Barrera}}, \bibinfo {author} {\bibfnamefont {R.}~\bibnamefont {Moyard}}, \bibinfo {author} {\bibfnamefont {Z.}~\bibnamefont {Niu}}, \bibinfo {author} {\bibfnamefont {L.~J.}\ \bibnamefont {O'Riordan}}, \bibinfo {author} {\bibfnamefont {S.}~\bibnamefont {Oud}}, \bibinfo {author} {\bibfnamefont {A.}~\bibnamefont {Panigrahi}}, \bibinfo {author} {\bibfnamefont {C.-Y.}\ \bibnamefont {Park}}, \bibinfo {author} {\bibfnamefont {D.}~\bibnamefont {Polatajko}}, \bibinfo {author} {\bibfnamefont {N.}~\bibnamefont {Quesada}}, \bibinfo {author} {\bibfnamefont {C.}~\bibnamefont {Roberts}},
  \bibinfo {author} {\bibfnamefont {N.}~\bibnamefont {Sá}}, \bibinfo {author} {\bibfnamefont {I.}~\bibnamefont {Schoch}}, \bibinfo {author} {\bibfnamefont {B.}~\bibnamefont {Shi}}, \bibinfo {author} {\bibfnamefont {S.}~\bibnamefont {Shu}}, \bibinfo {author} {\bibfnamefont {S.}~\bibnamefont {Sim}}, \bibinfo {author} {\bibfnamefont {A.}~\bibnamefont {Singh}}, \bibinfo {author} {\bibfnamefont {I.}~\bibnamefont {Strandberg}}, \bibinfo {author} {\bibfnamefont {J.}~\bibnamefont {Soni}}, \bibinfo {author} {\bibfnamefont {A.}~\bibnamefont {Száva}}, \bibinfo {author} {\bibfnamefont {S.}~\bibnamefont {Thabet}}, \bibinfo {author} {\bibfnamefont {R.~A.}\ \bibnamefont {Vargas-Hernández}}, \bibinfo {author} {\bibfnamefont {T.}~\bibnamefont {Vincent}}, \bibinfo {author} {\bibfnamefont {N.}~\bibnamefont {Vitucci}}, \bibinfo {author} {\bibfnamefont {M.}~\bibnamefont {Weber}}, \bibinfo {author} {\bibfnamefont {D.}~\bibnamefont {Wierichs}}, \bibinfo {author} {\bibfnamefont {R.}~\bibnamefont {Wiersema}}, \bibinfo {author}
  {\bibfnamefont {M.}~\bibnamefont {Willmann}}, \bibinfo {author} {\bibfnamefont {V.}~\bibnamefont {Wong}}, \bibinfo {author} {\bibfnamefont {S.}~\bibnamefont {Zhang}},\ and\ \bibinfo {author} {\bibfnamefont {N.}~\bibnamefont {Killoran}},\ }\href {https://arxiv.org/abs/1811.04968} {\bibinfo {title} {Pennylane: Automatic differentiation of hybrid quantum-classical computations}} (\bibinfo {year} {2022}),\ \Eprint {https://arxiv.org/abs/1811.04968} {arXiv:1811.04968 [quant-ph]} \BibitemShut {NoStop}%
\bibitem [{\citenamefont {Amari}(1993)}]{amari1993backpropagation}%
  \BibitemOpen
  \bibfield  {author} {\bibinfo {author} {\bibfnamefont {S.-i.}\ \bibnamefont {Amari}},\ }\bibfield  {title} {\bibinfo {title} {Backpropagation and stochastic gradient descent method},\ }\href {https://doi.org/10.1016/0925-2312(93)90006-O} {\bibfield  {journal} {\bibinfo  {journal} {Neurocomputing}\ }\textbf {\bibinfo {volume} {5}},\ \bibinfo {pages} {185} (\bibinfo {year} {1993})}\BibitemShut {NoStop}%
\bibitem [{\citenamefont {Preskill}(2018)}]{preskill2018quantum}%
  \BibitemOpen
  \bibfield  {author} {\bibinfo {author} {\bibfnamefont {J.}~\bibnamefont {Preskill}},\ }\bibfield  {title} {\bibinfo {title} {Quantum computing in the nisq era and beyond},\ }\href {https://doi.org/10.22331/q-2018-08-06-79} {\bibfield  {journal} {\bibinfo  {journal} {Quantum}\ }\textbf {\bibinfo {volume} {2}},\ \bibinfo {pages} {79} (\bibinfo {year} {2018})}\BibitemShut {NoStop}%
\bibitem [{\citenamefont {Cai}\ \emph {et~al.}(2023)\citenamefont {Cai}, \citenamefont {Babbush}, \citenamefont {Benjamin}, \citenamefont {Endo}, \citenamefont {Huggins}, \citenamefont {Li}, \citenamefont {McClean},\ and\ \citenamefont {O’Brien}}]{cai2023quantum}%
  \BibitemOpen
  \bibfield  {author} {\bibinfo {author} {\bibfnamefont {Z.}~\bibnamefont {Cai}}, \bibinfo {author} {\bibfnamefont {R.}~\bibnamefont {Babbush}}, \bibinfo {author} {\bibfnamefont {S.~C.}\ \bibnamefont {Benjamin}}, \bibinfo {author} {\bibfnamefont {S.}~\bibnamefont {Endo}}, \bibinfo {author} {\bibfnamefont {W.~J.}\ \bibnamefont {Huggins}}, \bibinfo {author} {\bibfnamefont {Y.}~\bibnamefont {Li}}, \bibinfo {author} {\bibfnamefont {J.~R.}\ \bibnamefont {McClean}},\ and\ \bibinfo {author} {\bibfnamefont {T.~E.}\ \bibnamefont {O’Brien}},\ }\bibfield  {title} {\bibinfo {title} {Quantum error mitigation},\ }\href {https://doi.org/10.1103/RevModPhys.95.045005} {\bibfield  {journal} {\bibinfo  {journal} {Rev. Mod. Phys.}\ }\textbf {\bibinfo {volume} {95}},\ \bibinfo {pages} {045005} (\bibinfo {year} {2023})}\BibitemShut {NoStop}%
\end{thebibliography}%

\end{document}